\documentclass[twocolumn]{aastex62}
\usepackage{longtable}
\usepackage{comment}
\graphicspath{{./}{figures/}}

\submitjournal{AAS Journals}

\shorttitle{TOI-481b \& TOI-892b}
\shortauthors{Brahm et al. 2020}

\begin{document}

\title{TOI-481 b \& TOI-892 b: Two long period hot Jupiters from the Transiting Exoplanet Survey Satellite}

\newcommand{\feh}{\ensuremath{{\rm [Fe/H]}}}
\newcommand{\teff}{\ensuremath{T_{\rm eff}}}
\newcommand{\teq}{\ensuremath{T_{\rm eq}}}
\newcommand{\logg}{\ensuremath{\log{g}}}
\newcommand{\zaspe}{\texttt{ZASPE}}
\newcommand{\ceres}{\texttt{CERES}}
\newcommand{\tess}{\textit{TESS}}
\newcommand{\vsini}{\ensuremath{v \sin{i}}}
\newcommand{\kms}{\ensuremath{{\rm km\,s^{-1}}}}
\newcommand{\mjup}{\ensuremath{{\rm M_{J}}}}
\newcommand{\mearth}{\ensuremath{{\rm M}_{\oplus}}}
\newcommand{\mpl}{\ensuremath{{\rm M_P}}}
\newcommand{\rjup}{\ensuremath{{\rm R_J}}}
\newcommand{\rpl}{\ensuremath{{\rm R_P}}}
\newcommand{\rstar}{\ensuremath{{\rm R}_{\star}}}
\newcommand{\mstar}{\ensuremath{{\rm M}_{\star}}}
\newcommand{\lstar}{\ensuremath{{\rm L}_{\star}}}
\newcommand{\rsun}{\ensuremath{{\rm R}_{\odot}}}
\newcommand{\msun}{\ensuremath{{\rm M}_{\odot}}}
\newcommand{\lsun}{\ensuremath{{\rm L}_{\odot}}}

\newcommand{\teffA}{\ensuremath{5735 \pm 72}}
\newcommand{\fehA}{\ensuremath{+0.26\pm 0.05}}
\newcommand{\vsiniA}{\ensuremath{4.54\pm 0.3}}
\newcommand{\ageA}{\ensuremath{6.7_{-0.6}^{+0.4}}}
\newcommand{\mstA}{\ensuremath{1.14_{-0.01}^{+0.02}}}
\newcommand{\loggA}{\ensuremath{4.06_{-0.01}^{+0.01}}}
\newcommand{\rstA}{\ensuremath{1.66_{-0.02}^{+0.02}}}
\newcommand{\lstA}{\ensuremath{2.72_{-0.08}^{+0.10}}}
\newcommand{\rhostA}{\ensuremath{0.36_{-0.01}^{+0.01}}}
\newcommand{\avA}{\ensuremath{0.06_{-0.04}^{+0.05}}}
\newcommand{\rpA}{\ensuremath{0.99_{-0.01}^{+0.01}}}
\newcommand{\mpA}{\ensuremath{1.53_{-0.03}^{+0.03}}}
\newcommand{\tqA}{\ensuremath{1370_{-10}^{+10}}}
\newcommand{\smaA}{\ensuremath{0.097_{-0.001}^{+0.001}}}
\newcommand{\mcoreA}{\ensuremath{76.57 \pm 7.71}}

\newcommand{\teffB}{\ensuremath{6261 \pm 80}}
\newcommand{\fehB}{\ensuremath{+0.24 \pm 0.05}}
\newcommand{\vsiniB}{\ensuremath{7.69 \pm 0.5}}
\newcommand{\ageB}{\ensuremath{2.2_{-0.5}^{+0.5}}}
\newcommand{\mstB}{\ensuremath{1.28_{-0.02}^{+0.03}}}
\newcommand{\loggB}{\ensuremath{4.26_{-0.02}^{+0.02}}}
\newcommand{\rstB}{\ensuremath{1.39_{-0.02}^{+0.02}}}
\newcommand{\lstB}{\ensuremath{2.7_{-0.1}^{+0.1}}}
\newcommand{\rhostB}{\ensuremath{0.67_{-0.03}^{+0.04}}}
\newcommand{\avB}{\ensuremath{0.18_{-0.08}^{+0.07}}}
\newcommand{\rpB}{\ensuremath{1.07_{-0.02}^{+0.02}}}
\newcommand{\mpB}{\ensuremath{0.95_{-0.07}^{+0.07}}}
\newcommand{\tqB}{\ensuremath{1397_{-40}^{+40}}}
\newcommand{\smaB}{\ensuremath{0.092_{-0.005}^{+0.005}}}
\newcommand{\mcoreB}{\ensuremath{25.04 \pm 9.14}}

\newcommand{\plnameA}{TOI-481~b}
\newcommand{\stnameA}{TOI-481}
\newcommand{\plnameB}{TOI-892~b}
\newcommand{\stnameB}{TOI-892}

\newcommand{\rhopl}{\ensuremath{{\rm \rho_P}}}
\newcommand{\gccm}{\ensuremath{\mathrm{g}\,\mathrm{cm}^{-3}}}

\newcommand{\PA}{\ensuremath{10.33111^{+0.00002}_{-0.00002}}}
\newcommand{\aA}{\ensuremath{12.52^{+0.03}_{-0.04}}}
\newcommand{\tcA}{\ensuremath{2458511.6418^{+0.0002}_{-0.0002}}}
\newcommand{\raA}{\ensuremath{0.32^{+0.04}_{-0.04}}}
\newcommand{\rbA}{\ensuremath{0.1228^{+0.0005}_{-0.0004}}}
\newcommand{\KA}{\ensuremath{130.3^{+1.4}_{-1.4}}}
\newcommand{\sesinomegaA}{\ensuremath{0.354^{+0.009}_{-0.010}}}
\newcommand{\secosomegaA}{\ensuremath{0.17^{+0.01}_{-0.01}}}

\newcommand{\qaTESSA}{\ensuremath{0.64^{+0.06}_{-0.06}}}
\newcommand{\qbTESSA}{\ensuremath{0.11^{+0.03}_{-0.03}}}
\newcommand{\qaCHATA}{\ensuremath{0.55^{+0.08}_{-0.08}}}
\newcommand{\mfluxCHATA}{\ensuremath{-0.0019^{+0.0001}_{-0.0002}}}
\newcommand{\sigmaCHATA}{\ensuremath{2815^{+95}_{-96}}}
\newcommand{\qaNGTSA}{\ensuremath{0.96^{+0.003}_{-0.04}}}
\newcommand{\mfluxNGTSA}{\ensuremath{0.00004^{+0.000007}_{-0.00007}}}

\newcommand{\muFEROSA}{\ensuremath{37797.1^{+1.7}_{-1.6}}}
\newcommand{\sigmawFEROSA}{\ensuremath{4.1^{+1.7}_{-1.3}}}
\newcommand{\muCHIRONA}{\ensuremath{41.1^{+1.4}_{-1.3}}}
\newcommand{\sigmawCHIRONA}{\ensuremath{0.018^{+0.02}_{-0.006}}}
\newcommand{\muAMinervaA}{\ensuremath{-54.5^{+3.4}_{-3.1}}}
\newcommand{\sigmawAMinervaA}{\ensuremath{14.7^{+3.5}_{-2.5}}}
\newcommand{\muBMinervaA}{\ensuremath{27.1^{+2.6}_{-2.5}}}
\newcommand{\sigmawBMinervaA}{\ensuremath{22.4^{+2.3}_{-1.9}}}
\newcommand{\muNRESA}{\ensuremath{-0.7^{+6.7}_{-6.8}}}
\newcommand{\sigmawNRESA}{\ensuremath{33^{+10}_{-10}}}
\newcommand{\muCORALIEA}{\ensuremath{37808.2^{+2.2}_{-2.2}}}
\newcommand{\muFIDEOSA}{\ensuremath{37086^{+16}_{-17}}}

\newcommand{\pA}{\ensuremath{0.0614^{+0.0002}_{-0.0002}}}
\newcommand{\bA}{\ensuremath{0.15^{+0.05}_{-0.05}}}
\newcommand{\incA}{\ensuremath{89.2^{+0.3}_{-0.3}}}
\newcommand{\eccA}{\ensuremath{0.153^{+0.006}_{-0.007}}}
\newcommand{\omegaA}{\ensuremath{64.8^{+1.8}_{-1.8}}}
\newcommand{\rhoA}{\ensuremath{347.8^{+2.8}_{-3.5}}}

\newcommand{\tcB}{\ensuremath{2458475.689^{+0.002}_{-0.002}}}
\newcommand{\PB}{\ensuremath{10.62656^{+0.00007}_{-0.00007}}}
\newcommand{\pB}{\ensuremath{0.079^{+0.001}_{-0.001}}}
\newcommand{\bB}{\ensuremath{0.43^{+0.09}_{-0.13}}}
\newcommand{\aB}{\ensuremath{14.2^{+0.8}_{-0.7}}}
\newcommand{\KB}{\ensuremath{0.074^{+0.005}_{-0.005}}}

\newcommand{\sigmawTESSERACTTESSB}{\ensuremath{553^{+20}_{-21}}}
\newcommand{\qaTESSERACTTESSB}{\ensuremath{0.4^{+0.2}_{-0.2}}}
\newcommand{\qbTESSERACTTESSB}{\ensuremath{0.4^{+0.3}_{-0.3}}}
\newcommand{\GPrhoTESSERACTTESSB}{\ensuremath{0.6^{+0.2}_{-0.2}}}
\newcommand{\GPsigmaTESSERACTTESSB}{\ensuremath{363^{+70}_{-51}}}

\newcommand{\sigmawCHATB}{\ensuremath{1046^{+69}_{-70}}}
\newcommand{\qaCHATB}{\ensuremath{0.6^{+0.2}_{-0.2}}}
\newcommand{\sigmawMEARTHB}{\ensuremath{869^{+77}_{-77}}}
\newcommand{\qaMEARTHB}{\ensuremath{0.90^{+0.08}_{-0.13}}}
\newcommand{\mfluxMEARTHBB}{\ensuremath{-0.0019^{+0.0001}_{-0.0001}}}
\newcommand{\muFEROSB}{\ensuremath{42.033^{+0.005}_{-0.005}}}
\newcommand{\sigmawFEROSB}{\ensuremath{0.012^{+0.006}_{-0.006}}}
\newcommand{\muTRESB}{\ensuremath{0.05^{+0.01}_{-0.01}}}
\newcommand{\sigmawTRESB}{\ensuremath{0.005^{+0.010}_{-0.003}}}

\newcommand{\rhoB}{\ensuremath{482^{+82}_{-72}}}

\correspondingauthor{Rafael Brahm} 
\email{rafael.brahm@uai.cl}
\author[0000-0002-9158-7315]{Rafael Brahm}
\affiliation{Facultad de Ingeniería y Ciencias, Universidad Adolfo Ib\'a\~nez, Av.\ Diagonal las Torres 2640, Pe\~nalol\'en, Santiago, Chile}
\affiliation{Millennium Institute for Astrophysics, Chile}
\author[0000-0002-5254-2499]{Louise D. Nielsen} 
\affiliation{Geneva Observatory, University of Geneva, Chemin des Maillettes 51, 1290 Versoix, Switzerland}
\author[0000-0001-9957-9304]{Robert A. Wittenmyer} 
\affiliation{University of Southern Queensland, Centre for Astrophysics, Toowoomba, QLD 4350, Australia}
\author[0000-0002-7846-6981]{Songhu Wang} 
\affiliation{Department of Astronomy, Yale University, New Haven, CT 06511, USA}
\author[0000-0001-8812-0565]{Joseph E. Rodriguez} 
\affiliation{Center for Astrophysics \textbar \ Harvard \& Smithsonian, 60 Garden St, Cambridge, MA 02138, USA}
\author[0000-0001-9513-1449]{N\'estor Espinoza} 
\affiliation{Space Telescope Science Institute, 3700 San Martin Drive, Baltimore, MD 21218, USA}
\author{Mat\'ias I. Jones} 
\affiliation{European Southern Observatory, Casilla 19001, Santiago, Chile}
\author[0000-0002-5389-3944]{Andr\'es Jord\'an} 
\affiliation{Facultad de Ingeniería y Ciencias, Universidad Adolfo Ib\'a\~nez, Av.\ Diagonal las Torres 2640, Pe\~nalol\'en, Santiago, Chile}
\affiliation{Millennium Institute for Astrophysics, Chile}
\author{Thomas Henning} 
\affiliation{Max-Planck-Institut f\"ur Astronomie, K\"onigstuhl 17, Heidelberg 69117, Germany }
\author[0000-0002-5945-7975]{Melissa Hobson} 
\affiliation{Millennium Institute for Astrophysics, Chile}
\affiliation{Instituto de Astrof\'isica, Pontificia Universidad Cat\'olica de Chile, Av.\ Vicu\~na Mackenna 4860, Macul, Santiago, Chile}
\author[0000-0002-0436-7833]{Diana Kossakowski} 
\affiliation{Max-Planck-Institut f\"ur Astronomie, K\"onigstuhl 17, Heidelberg 69117, Germany }
\author{Felipe Rojas} 
\affiliation{Instituto de Astrof\'isica, Pontificia Universidad Cat\'olica de Chile, Av.\ Vicu\~na Mackenna 4860, Macul, Santiago, Chile}
\affiliation{Millennium Institute for Astrophysics, Chile}
\author{Paula Sarkis} 
\affiliation{Max-Planck-Institut f\"ur Astronomie, K\"onigstuhl 17, Heidelberg 69117, Germany }
\author[0000-0001-8355-2107]{Martin Schlecker} 
\affiliation{Max-Planck-Institut f\"ur Astronomie, K\"onigstuhl 17, Heidelberg 69117, Germany }
\author{Trifon Trifonov} 
\affiliation{Max-Planck-Institut f\"ur Astronomie, K\"onigstuhl 17, Heidelberg 69117, Germany }
\author{Sahar Shahaf} 
\affiliation{School of Physics and Astronomy, Tel-Aviv University, Tel Aviv 69978, Israel}
%
\author{George Ricker} 
\affiliation{Department of Physics and Kavli Institute for Astrophysics and Space Research, Massachusetts Institute of Technology, Cambridge, MA 02139, USA}
\author{Roland Vanderspek} 
\affiliation{Department of Physics and Kavli Institute for Astrophysics and Space Research, Massachusetts Institute of Technology, Cambridge, MA 02139, USA}
\author{David W. Latham}
\affiliation{Center for Astrophysics \textbar \ Harvard \& Smithsonian, 60 Garden St, Cambridge, MA 02138, USA}
\author[0000-0002-6892-6948]{Sara Seager} 
\affiliation{Department of Physics and Kavli Institute for Astrophysics and Space Research, Massachusetts Institute of Technology, Cambridge, MA 02139, USA}
\affiliation{Department of Earth, Atmospheric and Planetary Sciences, Massachusetts Institute of Technology, Cambridge, MA 02139, USA}
\affiliation{Department of Aeronautics and Astronautics, MIT, 77 Massachusetts Avenue, Cambridge, MA 02139, USA}
\author{Joshua N. Winn} 
\affiliation{Department of Astrophysical Sciences, Princeton University, NJ 08544, USA}
\author{Jon M. Jenkins} 
\affiliation{NASA Ames Research Center, Moffett Field, CA 94035, USA}
\author[0000-0003-3216-0626]{Brett C. Addison} 
\affil{University of Southern Queensland, Centre for Astrophysics, West Street, Toowoomba, QLD 4350 Australia}
\author[0000-0001-7204-6727]{G\'asp\'ar \'A.\ Bakos} 
\affiliation{Department of Astrophysical Sciences, Princeton University, NJ 08544, USA}
\altaffiliation{MTA Distinguished Guest Fellow, Konkoly Observatory, Hungary}
\author[0000-0002-0628-0088]{Waqas Bhatti} 
\affiliation{Department of Astrophysical Sciences, Princeton University, NJ 08544, USA}

%
\author[0000-0001-6023-1335]{Daniel Bayliss} 
\affiliation{Dept. of Physics, University of Warwick, Gibbet Hill Road, Coventry CV4 7AL, UK}
\author{Perry Berlind} 
\affiliation{Center for Astrophysics \textbar \ Harvard \& Smithsonian, 60 Garden St, Cambridge, MA 02138, USA}
\author[0000-0001-6637-5401]{Allyson Bieryla} 
\affiliation{Center for Astrophysics \textbar \ Harvard \& Smithsonian, 60 Garden St, Cambridge, MA 02138, USA}
\author{Francois Bouchy} 
\affiliation{Geneva Observatory, University of Geneva, Chemin des Maillettes 51, 1290 Versoix, Switzerland}
\author[0000-0003-2649-2288]{Brendan P. Bowler}  
\affil{Department of Astronomy, The University of Texas at Austin, TX 78712, USA}
\author{C\'{e}sar Brice\~{n}o} 
\affiliation{Cerro Tololo Inter-American Observatory, Casilla 603, 1700000, La Serena, Chile} 
\author[0000-0001-5062-0847]{Timothy M. Brown} 
\affiliation{University of Colorado/CASA, Boulder, CO 80309, USA}
\affiliation{Las Cumbres Observatory Global Telescope Network, Santa Barbara, CA 93117, USA}
\author{Edward M. Bryant} 
\affiliation{Dept. of Physics, University of Warwick, Gibbet Hill Road, Coventry CV4 7AL, UK}
\affiliation{Centre for Exoplanets and Habitability, University of Warwick, Gibbet Hill Road, Coventry CV4 7AL, UK}
\author[0000-0003-1963-9616]{Douglas A. Caldwell} 
\affiliation{SETI Institute, Mountain View, CA, 94043, USA}
\affiliation{SETI Institute/NASA Ames Research Center, Moffett Field, CA 94035, USA}
\author[0000-0002-9003-484X]{David Charbonneau} 
\affiliation{Center for Astrophysics \textbar \ Harvard \& Smithsonian, 60 Garden St, Cambridge, MA 02138, USA}
\author[0000-0001-6588-9574]{Karen A. Collins} 
\affiliation{Center for Astrophysics \textbar \ Harvard \& Smithsonian, 60 Garden St, Cambridge, MA 02138, USA}

%
\author[0000-0002-5070-8395]{Allen B. Davis} 
\affiliation{Department of Astronomy, Yale University, New Haven, CT 06511, USA}
\author[0000-0002-9789-5474]{Gilbert A. Esquerdo} 
\affiliation{Center for Astrophysics \textbar \ Harvard \& Smithsonian, 60 Garden St, Cambridge, MA 02138, USA}
\author[0000-0003-3504-5316]{Benjamin J. Fulton} 
\affiliation{NASA Exoplanet Science Institute / Caltech-IPAC, Pasadena, CA}
\author[0000-0002-5169-9427]{Natalia~M.~Guerrero} 
\affiliation{Department of Physics and Kavli Institute for Astrophysics and Space Research, Massachusetts Institute of Technology, Cambridge, MA 02139, USA}

%
\author{Christopher E. Henze} 
\affiliation{NASA Ames Research Center, Moffett Field, CA 94035, USA}
\author{Aleisha Hogan} 
\affiliation{School of Physics and Astronomy, University of Leicester, University Road, Leicester LE1 7RH, UK}
\author[0000-0002-1160-7970]{Jonathan Horner} 
\affil{University of Southern Queensland, Centre for Astrophysics, West Street, Toowoomba, QLD 4350 Australia}

\author[0000-0003-0918-7484]{Chelsea~ X.~Huang} 
\affiliation{Department of Physics and Kavli Institute for Astrophysics and Space Research, Massachusetts Institute of Technology, Cambridge, MA 02139, USA}
\author{Jonathan Irwin} 
\affiliation{Center for Astrophysics \textbar \ Harvard \& Smithsonian, 60 Garden St, Cambridge, MA 02138, USA}
\author[0000-0002-7084-0529]{Stephen R. Kane} 
\affil{Department of Earth and Planetary Sciences, University of California, Riverside, CA 92521, USA}
\author[0000-0003-0497-2651]{John Kielkopf} 
\affil{Department of Physics and Astronomy, University of Louisville, Louisville, KY 40292, USA}

%
\author[0000-0003-3654-1602]{Andrew W. Mann} 
\affiliation{Department of Physics and Astronomy, The University of North Carolina at Chapel Hill, Chapel Hill, NC 27599-3255, USA}
\author{Tsevi Mazeh} 
\affiliation{School of Physics and Astronomy, Raymond and Beverly Sackler Faculty of Exact Sciences, Tel Aviv University, Tel Aviv, 6997801, Israel}
\author[0000-0003-1631-4170]{James McCormac} 
\affiliation{Dept. of Physics, University of Warwick, Gibbet Hill Road, Coventry CV4 7AL, UK}
\author[0000-0001-5807-7893]{Curtis McCully} 
\affiliation{Las Cumbres Observatory Global Telescope Network, Santa Barbara, CA 93117, USA}
\affiliation{Department of Physics, University of California, Santa Barbara, CA 93106-9530, USA}
\author[0000-0002-7830-6822]{Matthew W. Mengel} 
\affil{University of Southern Queensland, Centre for Astrophysics, West Street, Toowoomba, QLD 4350 Australia}
\author[0000-0002-4510-2268]{Ismael Mireles} 
\affiliation{Department of Physics and Kavli Institute for Astrophysics and Space Research, Massachusetts Institute of Technology, Cambridge, MA 02139, USA}
\author[0000-0002-4876-8540]{Jack Okumura} 
\affil{University of Southern Queensland, Centre for Astrophysics, West Street, Toowoomba, QLD 4350 Australia}

%
\author{Peter Plavchan} 
\affil{George Mason University, 4400 University Drive MS 3F3, Fairfax, VA 22030, USA}
\author[0000-0002-8964-8377]{Samuel N. Quinn} 
\affiliation{Center for Astrophysics \textbar \ Harvard \& Smithsonian, 60 Garden St, Cambridge, MA 02138, USA}
\author[0000-0003-2935-7196]{Markus Rabus} 
\affiliation{Las Cumbres Observatory Global Telescope, 6740 Cortona Dr., Suite 102, Goleta, CA 93111, USA}
\affiliation{Department of Physics, University of California, Santa Barbara, CA 93106-9530, USA}
\author{Sophie Saesen} 
\affiliation{Geneva Observatory, University of Geneva, Chemin des Maillettes 51, 1290 Versoix, Switzerland}
\author{Joshua~E.~Schlieder} 
\affiliation{Exoplanets and Stellar Astrophysics Laboratory, Code 667, NASA Goddard Space Flight Center, Greenbelt, MD 20771, USA}
\author{Damien Segransan} 
\affiliation{Geneva Observatory, University of Geneva, Chemin des Maillettes 51, 1290 Versoix, Switzerland}
\author{Bernie Shiao} 
\affiliation{Space Telescope Science Institute, 3700 San Martin Drive, Baltimore, MD 21218, USA}
\author[0000-0002-1836-3120]{Avi Shporer} 
\affil{Department of Physics and Kavli Institute for Astrophysics and Space Research, Massachusetts Institute of Technology, Cambridge, MA 02139, USA}
\author[0000-0001-5016-3359]{Robert J. Siverd} 
\affiliation{Gemini Observatory/NSF’s NOIRLab, 670 N. A’ohoku Place, Hilo, HI, 96720, USA}
\author[0000-0002-3481-9052]{Keivan G. Stassun} 
\affiliation{Vanderbilt University, Department of Physics \& Astronomy, 6301 Stevenson Center Lane, Nashville, TN 37235, USA}
\author[0000-0001-7070-3842]{Vincent Suc} 
\affiliation{Facultad de Ingeniería y Ciencias, Universidad Adolfo Ib\'a\~nez, Av.\ Diagonal las Torres 2640, Pe\~nalol\'en, Santiago, Chile}
\affiliation{El Sauce Observatory, Chile}
\author[0000-0001-5603-6895]{Thiam-Guan Tan} 
\affiliation{Perth Exoplanet Survey Telescope, Perth, Australia}
\author{Pascal Torres} 
\affiliation{Instituto de Astrof\'isica, Pontificia Universidad Cat\'olica de Chile, Av.\ Vicu\~na Mackenna 4860, Macul, Santiago, Chile}
\affiliation{Millennium Institute for Astrophysics, Chile}
\author[0000-0002-7595-0970]{Chris G. Tinney} 
\affil{Exoplanetary Science at UNSW, School of Physics, UNSW Sydney, NSW 2052, Australia}
\author{Stephane Udry} 
\affiliation{Geneva Observatory, University of Geneva, Chemin des Maillettes 51, 1290 Versoix, Switzerland}
\author{Leonardo Vanzi} 
\affiliation{Department of Electrical Engineering, Pontificia Universidad Cat\'olica de Chile, Av. Vicu\~{n}a Mackenna 4860, 7820436 Macul, Santiago, Chile}
\affiliation{Center of Astro-Engineering UC, Pontificia Universidad Cat\'olica de Chile, Av. Vicu\~{n}a Mackenna 4860, 7820436 Macul, Santiago, Chile}
\author{Michael~Vezie} 
\affiliation{Department of Physics and Kavli Institute for Astrophysics and Space Research, Massachusetts Institute of Technology, Cambridge, MA 02139, USA}
\author[0000-0002-1896-2377]{Jose I. Vines} 
\affiliation{Departamento de Astronom\'ia, Universidad de Chile, Camino El Observatorio 1515, Las Condes, Santiago, Chile}
\author{Maja Vuckovic} 
\affiliation{Instituto de F\'isica y Astronom\'ia, Universidad de Vapara\'iso, Casilla 5030, Valpara\'iso, Chile}
\author[0000-0001-7294-5386]{Duncan J. Wright} 
\affil{University of Southern Queensland, Centre for Astrophysics, West Street, Toowoomba, QLD 4350 Australia}
\author[0000-0003-4755-584X]{Daniel A. Yahalomi}  
\affiliation{Center for Astrophysics \textbar \ Harvard \& Smithsonian, 60 Garden St, Cambridge, MA 02138, USA}
\author[0000-0003-2326-6488]{Abner Zapata} 
\affiliation{Department of Electrical Engineering, Pontificia Universidad Cat\'olica de Chile, Av. Vicu\~{n}a Mackenna 4860, 7820436 Macul, Santiago, Chile}
\affiliation{Center of Astro-Engineering UC, Pontificia Universidad Cat\'olica de Chile, Av. Vicu\~{n}a Mackenna 4860, 7820436 Macul, Santiago, Chile}
\affiliation{El Sauce Observatory, Chile}
\author[0000-0003-3491-6394]{Hui Zhang} 
\affil{School of Astronomy and Space Science, Key Laboratory of Modern Astronomy and Astrophysics in Ministry of Education, Nanjing University, Nanjing 210046, Jiangsu, China}
\author{Carl Ziegler} 
\affiliation{Dunlap Institute for Astronomy and Astrophysics, University of Toronto, 50 St. George Street, Toronto, Ontario M5S 3H4, Canada}



\begin{abstract}
We present the discovery of two new 10-day period giant planets from the Transiting Exoplanet Survey Satellite (\textit{TESS}) mission, whose masses were precisely determined using a wide diversity of ground-based facilities. \plnameA\ and \plnameB\ have similar radii ($0.99\pm0.01$ \rjup\ and $1.07\pm0.02$ \rjup, respectively), and orbital periods (10.3311 days and 10.6266 days, respectively), 
but significantly different masses ($1.53\pm0.03$ \mjup\ versus $0.95\pm0.07$ \mjup, respectively). Both planets orbit metal-rich stars (\feh\ = \fehA\ dex and \feh\ = \fehB\, for \stnameA\ and \stnameB, respectively) but at different evolutionary stages. \stnameA\ is a \mstar\ = $1.14\pm0.02$ \msun, \rstar\ = $1.66\pm0.02$ \rsun\ G-type star (\teff\ = \teffA\ K), that with an age of 6.7 Gyr, is in the turn-off point of the main sequence. \stnameB\, on the other hand, is a F-type dwarf star (\teff\ = \teffB\ K), which has a mass of \mstar\ = $1.28\pm0.03$ \msun, and a radius of \rstar\ = $1.39\pm0.02$ \rsun. \plnameA\ and \plnameB\ join the scarcely populated region of transiting gas giants with orbital periods longer than 10 days, which is important to constrain theories of the formation and structure of hot Jupiters.

\end{abstract}



\section{Introduction} \label{sec:intro}
Among the vast diversity of extrasolar planets discovered throughout the past three decades, those known as hot Jupiters \citep[e.g.,][]{mayor:1995} are arguably the most well-studied population. These objects are gas giant planets (\rpl\ $\gtrsim$ 0.8 \rjup) orbiting closely around their host stars, with typical orbital periods shorter than $\approx$10 days.

Despite having a relatively low occurrence rate of $\approx$1\% \citep{wang:2015,zhou:2019}, due to strong observational biases favoring their detection and characterization, hot Jupiters represent $\approx$75\% of the total sample of transiting extrasolar planets for which both masses and radii are determined with a precision of at least 20\% \footnote{based on the catalogue of the physical properties of transiting planetary systems \citep[TEPCat,][]{tepcat}, updated on July 7, 2020}.

Follow-up observations of hot Jupiters have delivered signifiant scientific results - including the first studies on the atmospheres of planets outside our own solar system \citep[e.g.,][]{charbonneau:2002,vidal:2003,pont:2008}; and significant misalignments between orbital and stellar spin axes  \citep{queloz:2010,winn:2010,hebrard:2011}.

While in principle the large amount of information available for transiting hot Jupiters should help us in unveiling the formation and evolution mechanisms that allow the existence of close-in gas giants, their extreme environments produced by the proximity to the host stars makes the interpretation of hot Jupiter properties a challenging task \citep[see][for a comprehensive review]{dawson:2018}. The exact formation/migration mechanism of hot Jupiters \citep[e.g.,][]{wu:2011,beauge:2012,naoz:2012}, and the mechanism responsible for generating highly inflated radii \citep[e.g.,][]{bodenheimer:2001,batygin:2010,leconte:2010,kurokawa:2015} are some of the active challenges in the field.

Gas giants with orbital periods longer than that of typical hot Jupiters (often called ``warm Jupiters'') should not be significantly influenced by these proximity effects, making the orbital and physical characterization of warm Jupiters an important step to solve some of the aforementioned challenges \citep[e.g.,][]{dong:2014,lopez:2016,thorngren:2016}. Ground-based photometric surveys \citep[e.g.,][]{bakos:2004,pollacco:2006,pepper:2007,bakos:2013}, which have discovered the vast majority ($\approx$80\%) of bright transiting hot Jupiter systems, have strong limitations for discovering planets with periods longer than P$\gtrsim$8 days \citep{gaudi:2005}. Space based missions such as \textit{Kepler} \citep{borucki:2010}, \textit{Kepler-K2} \citep{howell:2014}, and CoRoT \citep{corot} allowed the discovery and orbital characterization of the first two dozen of such systems \citep{bonomo:2010,deeg:2010,almenara:2018,shporer:2017,brahm:2018,jordan:2018}, but due to its significantly larger field of view, the \textit{TESS} mission \citep{tess} is expected to significantly increase that number \citep{sullivan:2015, barclay:2018}. In just its first two years of operation, \textit{TESS} has demonstrated its ability to discover transiting warm Jupiters suitable for characterization follow-up  \citep{nielsen,huber:2019,toi172,addison:2020,gill:2020}, and this number will grow with the extended mission \citep{cooke:2019}.

Here we present the discovery and orbital characterization of two gas giants located in the relatively sparsely populated parameter space of orbital periods slightly longer than 10 days. These discoveries were realized in the context of the Warm gIaNts with tEss (WINE) collaboration, which focuses on the systematic characterization of \textit{TESS} transiting warm giant planets \citep[e.g.,][]{hd1397,jordan:2020}.

The paper is structured as follows. In Section \ref{sec:obs} we present the \textit{TESS} data, and follow-up photometric and spectroscopic observations that allowed the discovery of both planets. In Section \ref{sec:anal} we describe the routines adopted to estimate the stellar parameters of both host stars and the final physical and orbital parameters of \plnameA\ and \plnameB. Our findings are discussed in Section \ref{sec:disc}.

\section{Observations} 
\label{sec:obs}

\subsection{TESS} 
\label{sec:tess}
\stnameA\ and \stnameB\  were monitored by \textit{TESS} during its first year of operation. \stnameA\ was observed in short cadence (2 minutes) mode in Sectors 6, 7, 9, 10, and 13, and in long cadence (30 minutes) mode in Sector 3. On the other hand, \stnameB\ was only observed in Sector 6, in long cadence mode. Transiting candidates were identified on both stars by the \textit{TESS} Science Office, and were released as \textit{TESS} Object of Interest (TOI) to the community. \plnameA\ was identified as a candidate based on two clear ``transit-like" features present in the SPOC light curve \citep{spoc} of Sector 6. \stnameA\ presented a strong detection at 68$\sigma$ and passed all the diagnostic tests conducted and presented in the Data Validation report \citep{twicken:2018,li:2019}, including the odd/even transit depth test, and the difference image centroiding and ghost diagnostic tests (which help reject false positives due to background sources). No additional transit-like signals were identified in the light curve.
On the other hand, \plnameB\ was reported as a \textit{TESS} alert on July 12, 2019 based on the analysis of the quick look pipeline \citep{huang:2019} of Sector 6. For both candidates the predicted planetary radii were consistent with being Jovian planets with orbital periods close to 10 days.

\begin{figure*}
\plotone{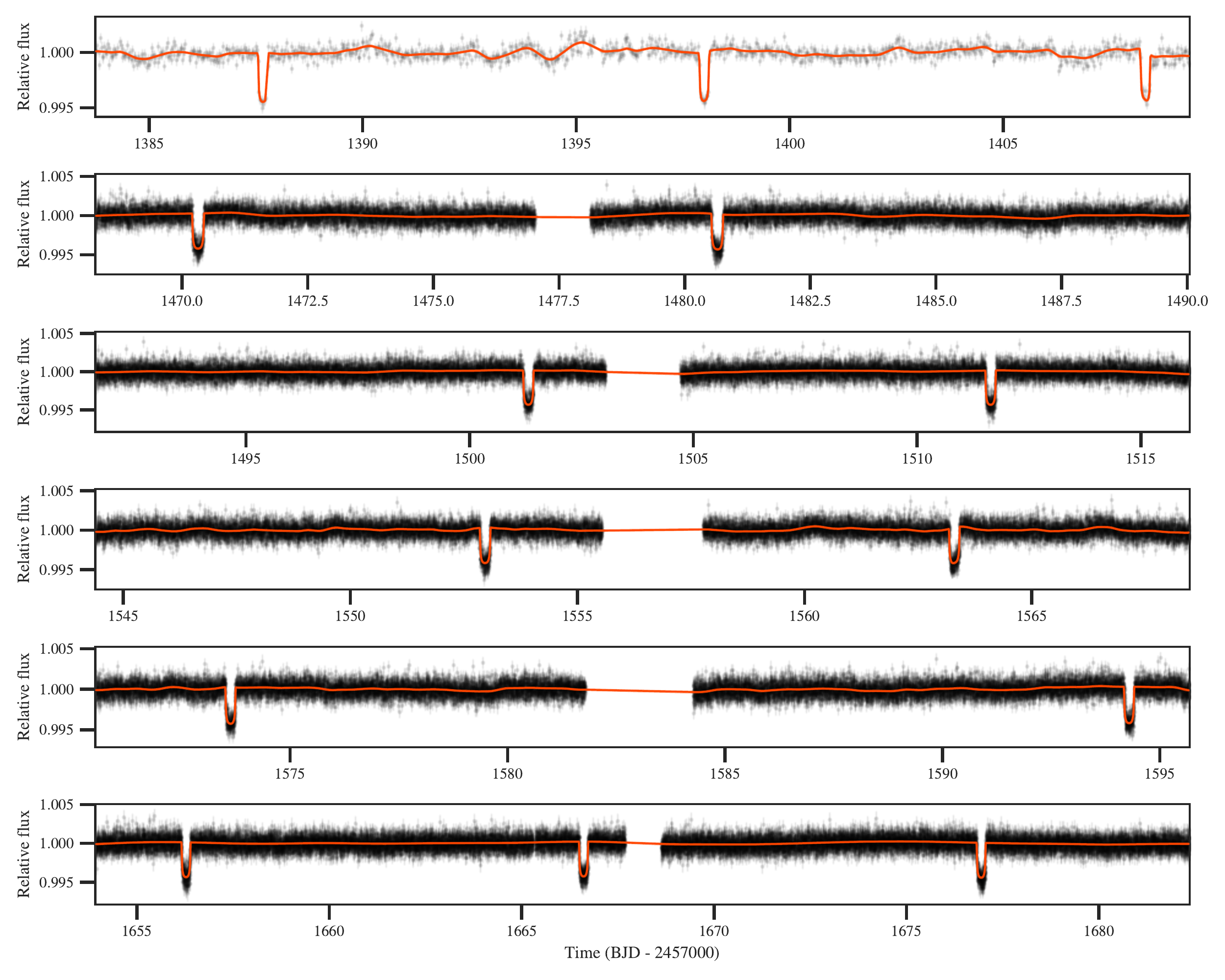}
\caption{\stnameA\ light curves of the six \textit{TESS} Sectors used in our analysis. The top panel presents the TESS Sector 3 data from the Full Frame Images as black points 
with errorbars obtained with \texttt{tesseract} (see text), while the rest of 
the panels show the 2-minute cadence Pre-search Data Conditioning Simple Aperture Photometry light curves for Sectors 6, 7, 9, 10 and 13, respectively. The orange line corresponds to the model obtained in Section \ref{sec:anal}, which consists of a transit model combined with a Gaussian process that describes the remaining flux variability.
}
\label{fig:tess_toi481}
\end{figure*}

\begin{figure*}
\plotone{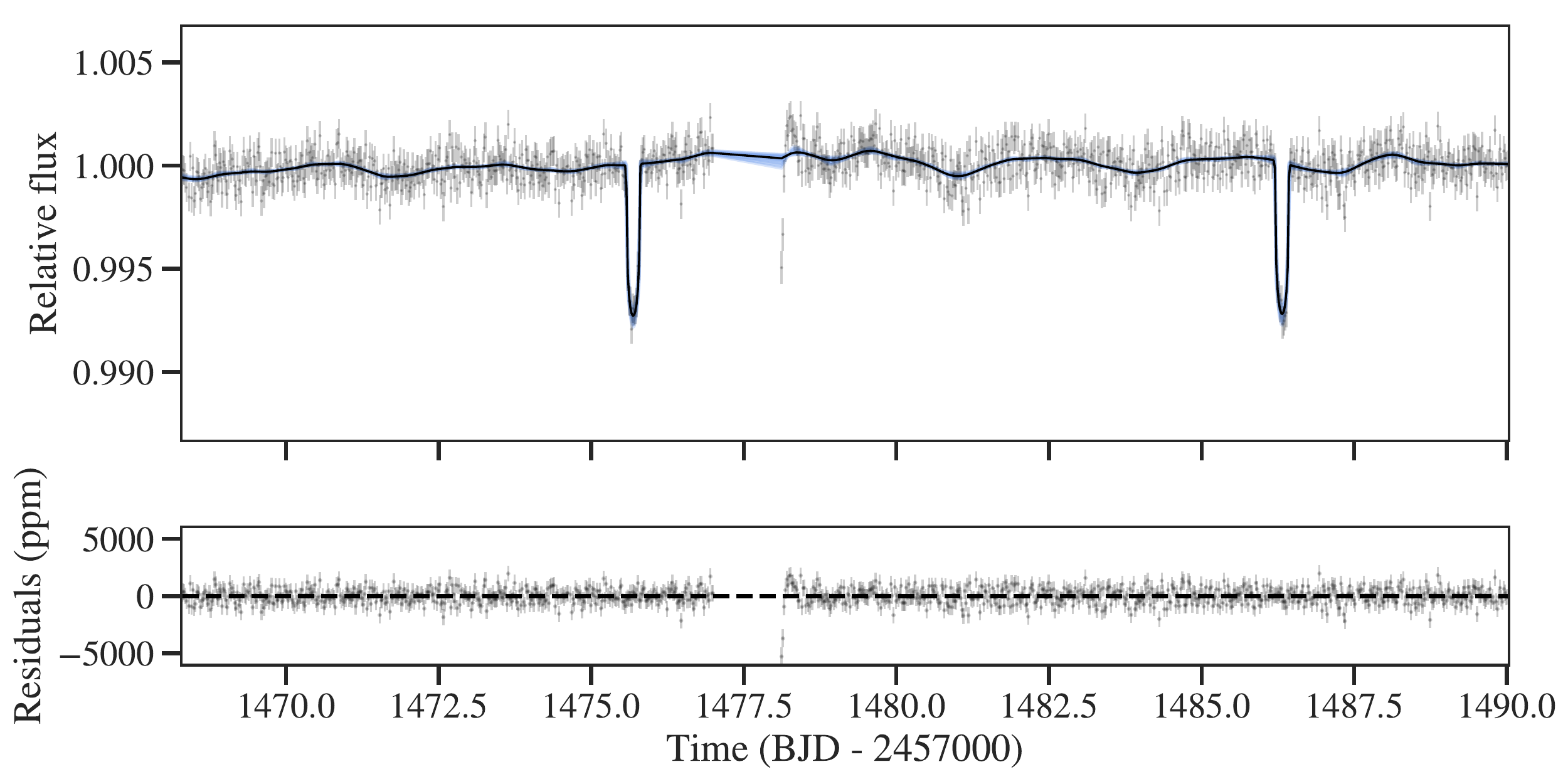}
\caption{The top panel corresponds to the TESS 30 minutes cadence light curve of \stnameB\ generated through \texttt{tesseract} from the Full Frame Images of Sector 6. The solid line corresponds to the model generated from the posterior parameters of the analysis presented in Section \ref{sec:anal}. The bottom panel shows the residuals between the \textit{TESS} light curve and the model.
}
\label{fig:phot_toi892}
\end{figure*}

For the \stnameA\ analysis presented in this study, we downloaded the Pre-search Data Conditioning Simple Aperture Photometry light curves \citep{stumpe:2012} of Sectors 6, 7, 9, 10, and 13 from the Mikulski Archives for Space Telescopes (see Figure \ref{fig:tess_toi481}). Systematic trends were removed from these light curves using the co-trending basis vectors \citep{smithPDC:2012,stumpePDC:2014}, generated by the \textit{TESS} SPOC at NASA Ames Research Center. We additionally obtained the long cadence light curve from the Full Frame Images of Sector 3 by using the \texttt{tesseract\footnote{\url{https://github.com/astrofelipe/tesseract}}} pipeline. For the analysis of \stnameB, we generated the long cadence light curve from the Full Frame Images of Sector 6 through \texttt{tesseract} (see Figure \ref{fig:phot_toi892}). The long cadence light curves for \stnameA\ and \stnameB\ used in this study are listed in Table \ref{tab:lc}.

\begin{deluxetable}{lrrrr}
\tablewidth{0pc}
\tablecaption{
    Long cadence (30 minutes) \textit{TESS} light curve data for \stnameA\ and \stnameB\ obtained from the \texttt{tesseract} extraction of the Full Frame Images of Sector 3 and 6, respectively . \label{tab:lc}
}
\tablehead{
    \colhead{ID} &
    \colhead{BJD} & 
    \colhead{Flux} & 
    \colhead{\ensuremath{\sigma_{\rm Flux}}} &
    \colhead{Sector} \\
}
\startdata
TOI-481 & 2458382.051879883 & 25279.4 & 4.8 & 3 \\
TOI-481 & 2458382.072692871 & 25272.1 & 4.7 & 3 \\
TOI-481 & 2458382.093566895 & 25274.6 & 4.7 & 3 \\
TOI-481 & 2458382.114379883 & 25282.0 & 4.7 & 3 \\
TOI-481 & 2458382.135192871 & 25279.4 & 4.7 & 3 \\
TOI-481 & 2458382.156066895 & 25277.6 & 4.7 & 3 \\
TOI-481 & 2458382.176940918 & 25283.0 & 4.7 & 3 \\
TOI-481 & 2458382.197753906 & 25282.9 & 4.7 & 3 \\
TOI-481 & 2458382.218566895 & 25277.7 & 4.7 & 3 \\
TOI-481 & 2458382.239440918 & 25299.6 & 4.7 & 3 \\
\enddata
\tablecomments{
    This table is available in a machine-readable form in the online
    journal.  A portion is shown here for guidance regarding its form
    and content.
}
\end{deluxetable}

\subsection{Ground-based photometry}
\label{sec:phot}
The limited spatial resolution of the \textit{TESS} mission and its relatively large pixel scale ($21\arcsec$/pix) makes necessary the execution of ground-based photometric observations to confirm that the transit features occur on target and not on close neighbor stars. Transits of both candidates were monitored with three different ground-based facilities installed in Chile. These observations were performed in the context of the \textit{TESS} Follow-up Observing Program (TFOP) Working Group Sub Group 1 (SG1). The four photometric timeseries are  publicly available on the Exoplanet Follow-up Observing Program for \textit{TESS} (ExoFOP-TESS) website\footnote{\url{https://exofop.ipac.caltech.edu/tess}}.

\subsubsection{CHAT}
The Chilean Hungarian Automated Telescope\footnote{\url{https://www.exoplanetscience2.org/sites/default/files/submission-attachments/poster_aj.pdf}} (CHAT) is a robotic facility installed at Las Campanas Observatory in Chile. CHAT consists in a FORNAX 200 equatorial mount, and a 0.7 m telescope coupled to a FLI ML-23042 CCD of 2048$\times$2048 pixels, which delivers a pixel scale of $0.6\arcsec$/pix. CHAT contains a set of \textit{i}', \textit{r}', and \textit{g}' passband filters.

\stnameA\ was observed with CHAT on the night of March 30, 2019 with the \textit{i}' filter adopting an exposure time of 20 s. We obtained 516 images of \stnameA\ with airmass values between 1.2 and 2. CHAT data were processed with a dedicated pipeline that performs differential aperture photometry, where the optimal comparison sources and the radius of the photometric aperture are automatically selected \citep[e.g.,][]{espinoza:chat,jones:2019,jordan:2018}. The light curve obtained is presented in the left panel of Figure \ref{fig:lcs_fold} and shows an ingress for \plnameA\ which confirms that the transit identified in \textit{TESS} data occurs in a region of $8\arcsec$ centered on \stnameA.

\stnameB\ was photometrically monitored with CHAT on the night of November 27, 2019. The \textit{i}' filter was used to obtain 189 images with an exposure time of 66 s. The right panel of Figure \ref{fig:lcs_fold} presents the CHAT light curve obtained for \plnameB, where a $\approx$7000 ppm egress can be identified, ensuring that the transit occurs inside $6\arcsec$ from \stnameB.





\subsubsection{MEarth-South}
The MEarth-South project \citep{irwin:2015} consists in an array of eight identical robotic 0.4 m telescopes installed in the Cerro Tololo International Observatory, in Chile. Seven telescopes of the array were used to monitor a transit of \plnameB\ the night of February 20, 2020. Each of the telescopes obtained approximately 360 images with a cadence of 52 s using a custom made RG715 filter. The data were processed with the MEarth South pipeline producing the light curve displayed in Figure \ref{fig:lcs_fold}, which further confirms the occurrence of the transit on target by registering an ingress.

\subsubsection{NGTS}
\begin{figure*}
\epsscale{1.1}
\plottwo{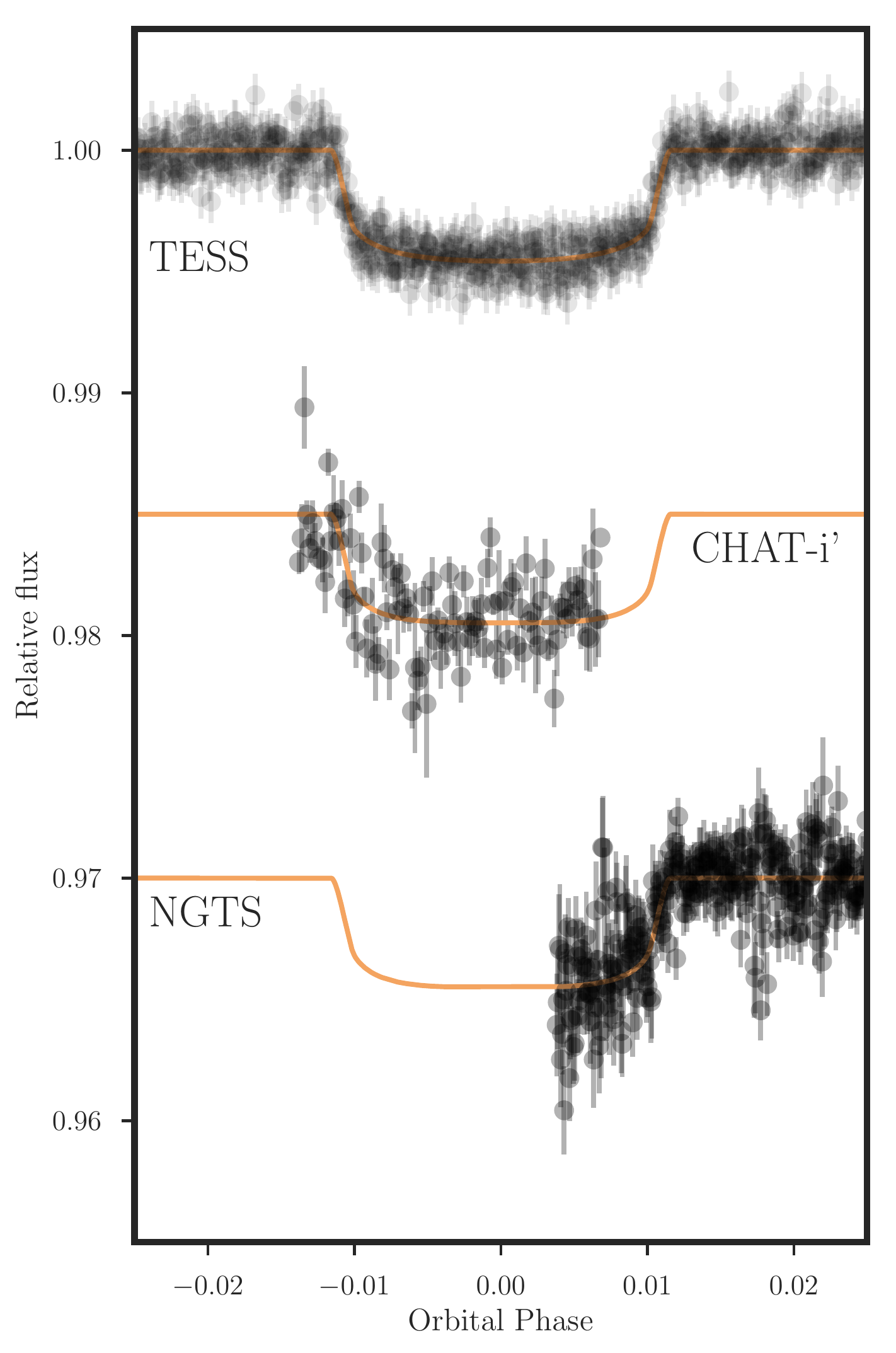}{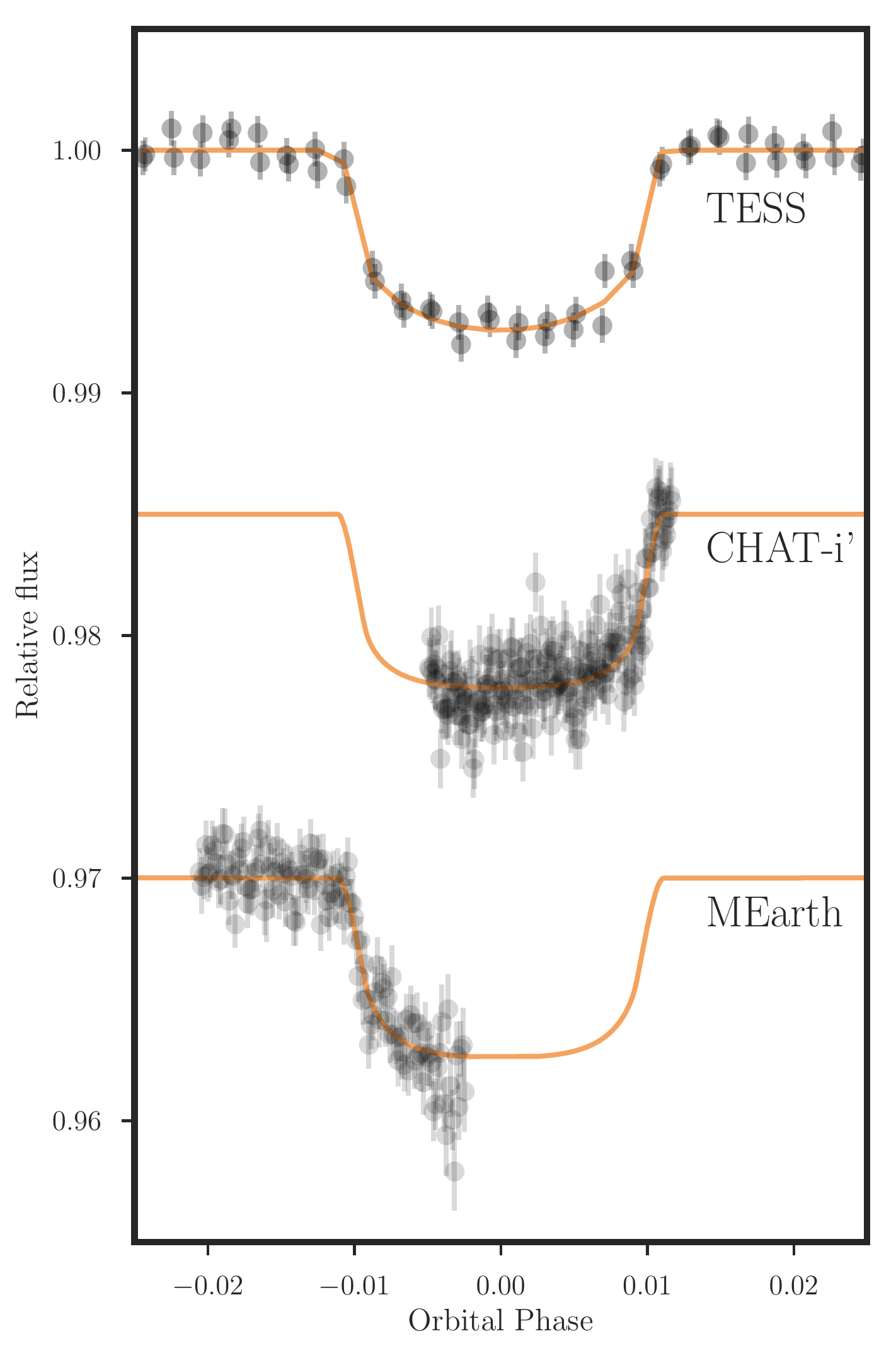}
\caption{The left panel shows the phase folded 2 minute cadence \tess\ photometry of \stnameA\ along with the ground-based follow-up light curves of CHAT and NGTS. The solid line shows the corresponding transit model in each case. The right panel shows the phase folded \textit{TESS} long cadence, CHAT, and MEarth light curves for \stnameB.} \label{fig:lcs_fold}
\end{figure*}

The Next Generation Transit Survey \citep[NGTS,][]{wheatley:2018} is an array of twelve identical robotic telescopes installed at the Paranal Observatory in Chile. Four NGTS telescopes were used simultaneously on the night of December 3, 2019 to monitor an egress of \plnameA.  Exposures were taken using a custom NGTS filter (520-890\,nm) with 10\,s exposure times which resulted in a $\sim$12\,s cadence.  Data were reduced using the NGTS aperture photometry pipeline detailed in \citet{bryant:2020}.  The NGTS light curve is presented in Figure \ref{fig:lcs_fold}.

\subsection{High resolution imaging}
\begin{figure*}
\plottwo{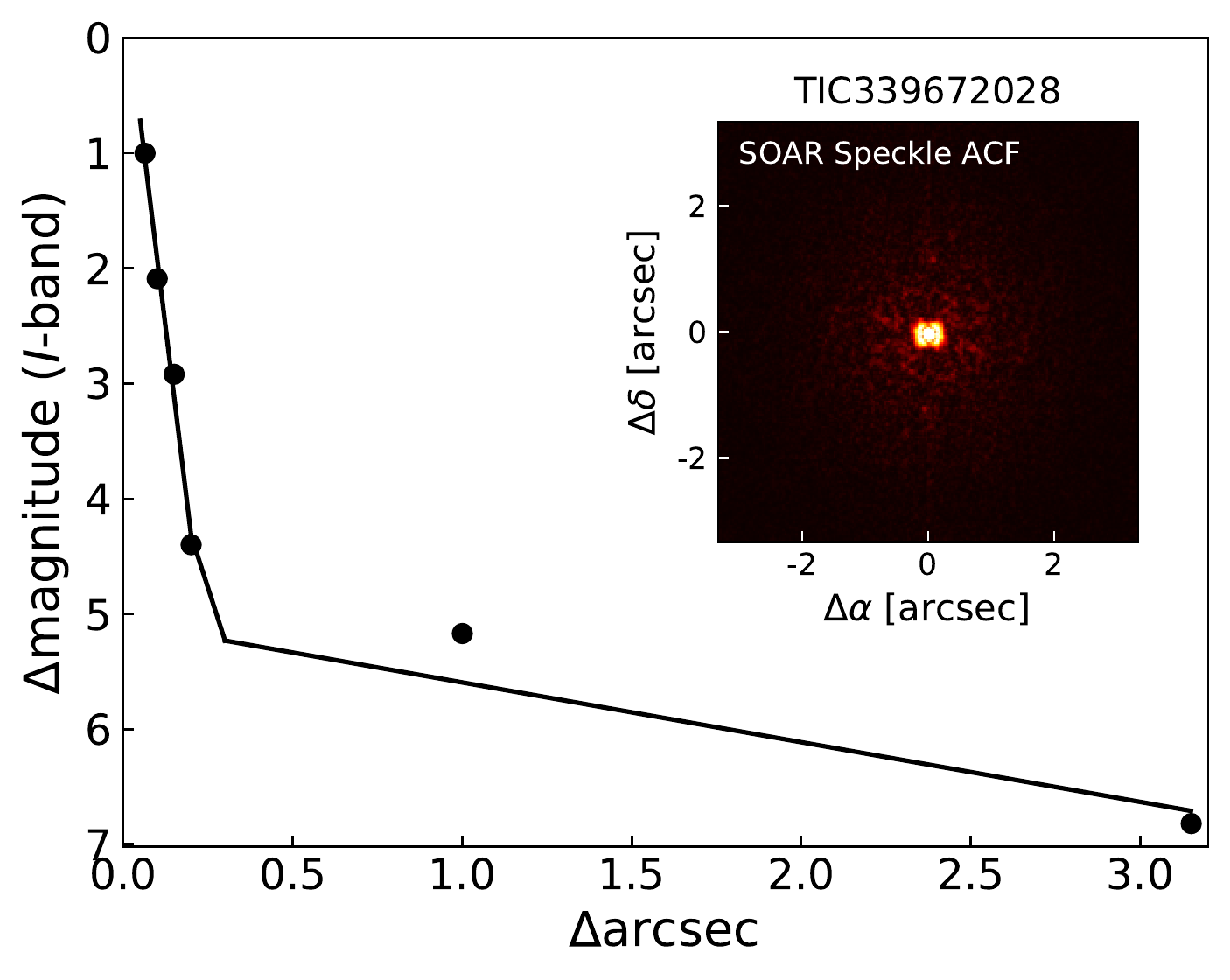}{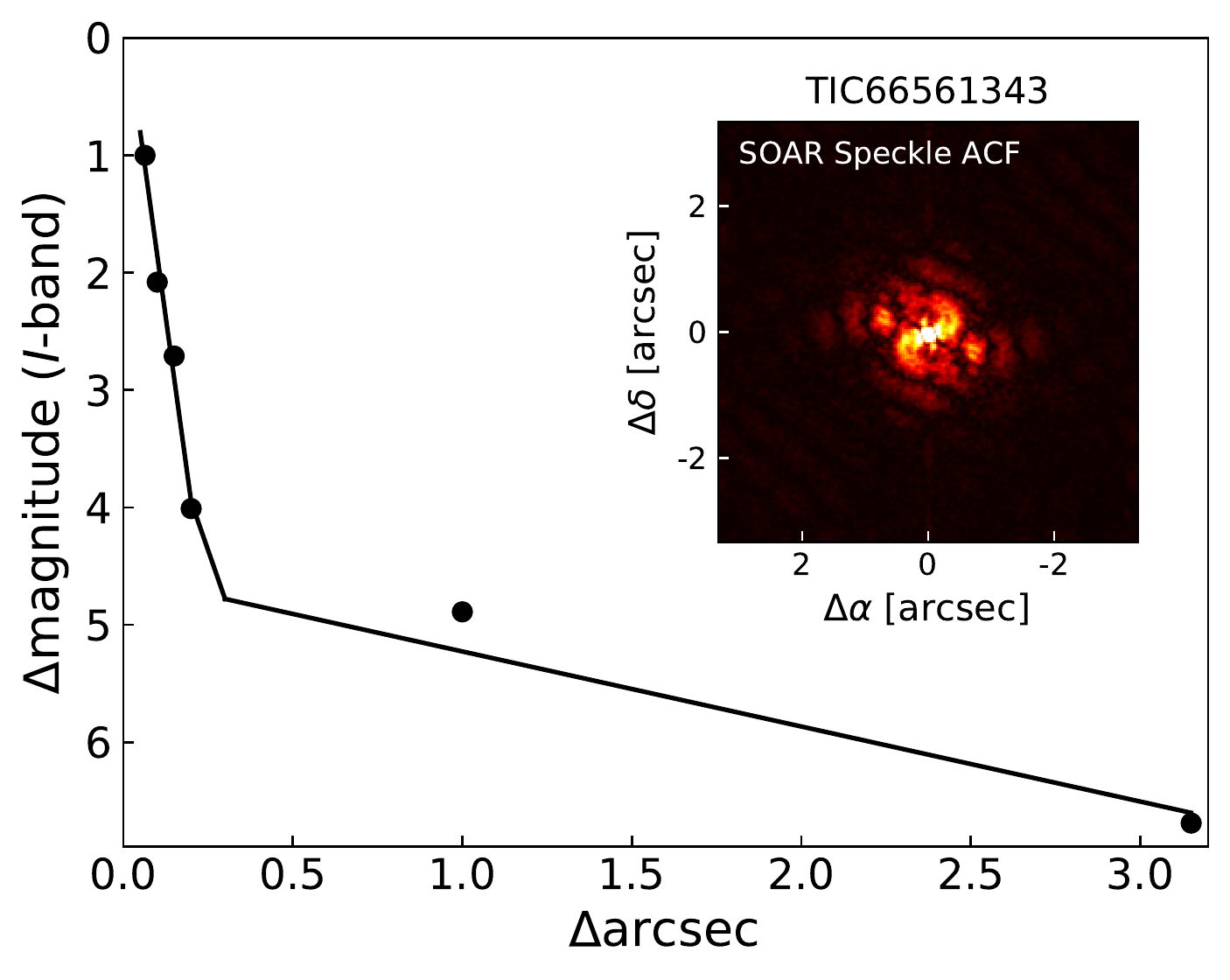}
\caption{
Contrast curve plots and auto-correlation functions from Speckle imaging in the I-band using the HRcam at SOAR, for \stnameA\ (left panel) and \stnameB\ (right panel). The black points correspond to the 5$\sigma$ contrast curve for each star. The solid line is the linear fit to the data for separations $<$0.2\arcsec and $>$0.2\arcsec.
}
\label{fig:soar}
\end{figure*}

The identification of contaminating sources in the neighborhood of transiting candidates is important for constraining false positive scenarios and for determining possible dilutions of the transits.
In this context, \stnameA\ and \stnameB\ were imaged with the High-Resolution Camera (HRCam) installed at the 4.1m Southern Astrophysical Research \citep[SOAR,][]{SOAR} telescope, in Cerro Pach\'on, Chile. Observations took place on the night of November 9, 2019, in the context of the SOAR \textit{TESS} Survey \citep{ziegler}. No nearby sources were detected in the vicinity of either star (see Figure \ref{fig:soar}).

We also used the Gaia DR2 catalog \citep{gaia:dr2} to identify the presence of close companions that could dilute the transit depths of \plnameA\ and \plnameB\ obtained from the ground-based light curves presented in section \ref{sec:phot}. We find that inside $10\arcsec$ from the target, \stnameA\ contains just one source having a magnitude difference of 8.7 mag in the G passband filter, which is too faint to significantly affect the transit depth of \plnameA. \stnameB\ reports no nearby sources closer than $10\arcsec$ to it. 

\subsection{High resolution spectroscopy} 
\label{sec:spec}
\begin{figure*}
\epsscale{2.3}
\plottwo{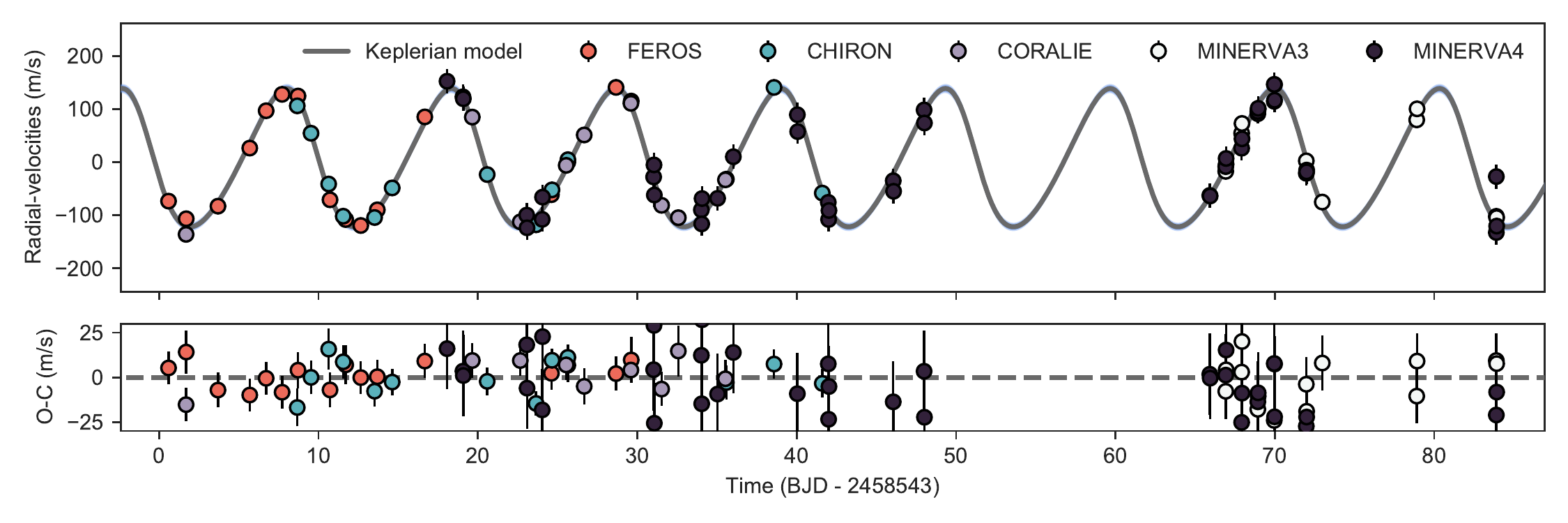}{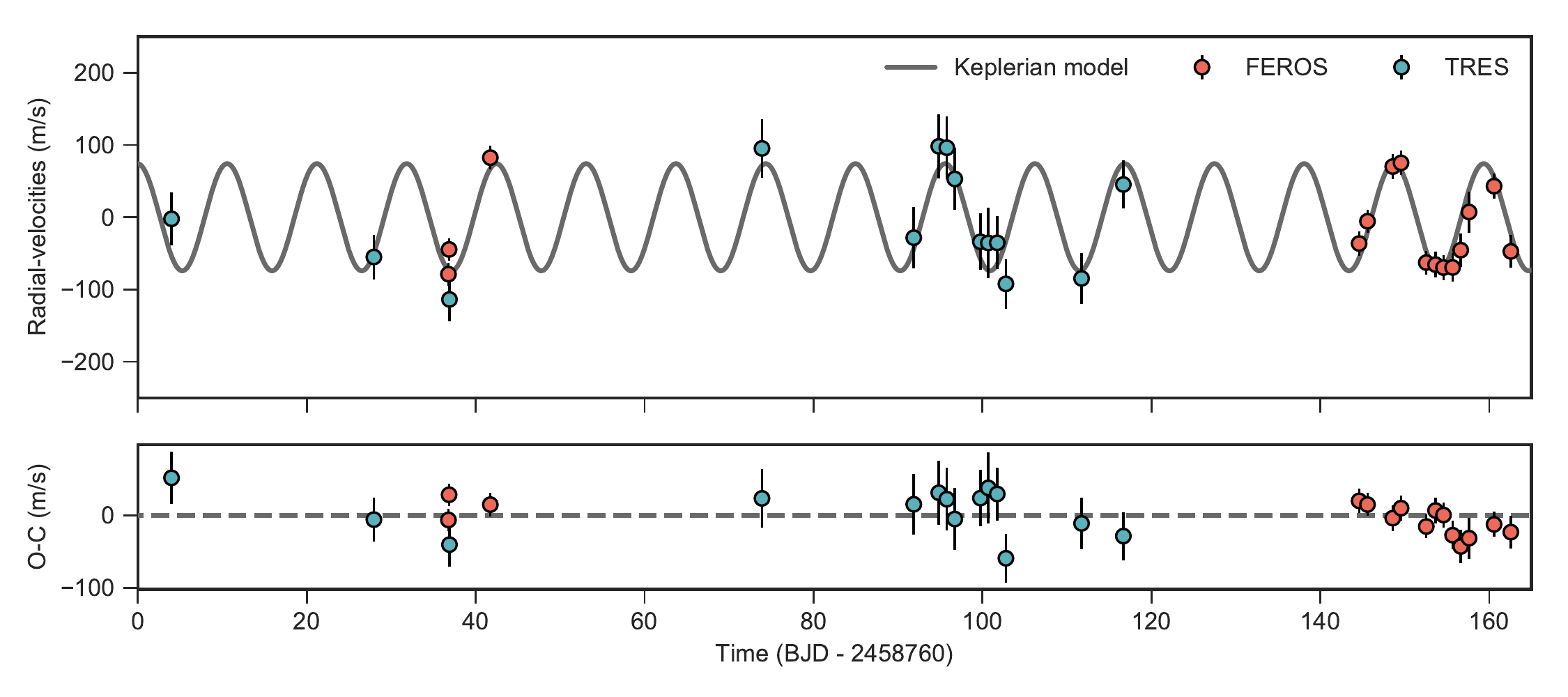}
\caption{Radial velocity observations for \stnameA\ (top panel) and \stnameB\ (bottom panel). The solid line corresponds to a Keplerian model using the posterior parameters of the global modeling presented in Section \ref{sec:anal}. The residuals are also presented below the radial velocity curves for each system. The radial velocity measurements for NRES have been excluded from the plots due to their significantly larger error bars.\label{fig:rvs}}
\end{figure*}

\stnameA\ and \stnameB\ were monitored with seven different spectrographs with the goal of measuring radial velocity variations to confirm the planetary nature of the transiting candidates and constrain their orbital parameters and masses. These observations are described in the following paragraphs and the radial velocities are presented in Table \ref{tab:rvs} and displayed in Figure \ref{fig:rvs}.

\subsubsection{FEROS}
The Fiberfed Extended Range Optical Spectrograph \citep[FEROS,][]{kaufer:99} has a resolving power of R=48,000 and is installed on the MPG2.2 m telescope at La Silla Observatory, in Chile. For this study, all FEROS data were processed with the \ceres\ pipeline \citep{brahm:2017:ceres}, which delivers optimally extracted, wavelength calibrated, and instrumental drift corrected spectra, along with the radial velocity and bisector span measurements.

We obtained 16 spectra with FEROS of \stnameA\ over a time span of 30 days starting on the night of February 28, 2019. We adopted an exposure time of 300 s which generated spectra with a signal-to-noise ratio per resolution element of $\sim$110. 

For \stnameB\ we obtained 15 FEROS spectra between November 9, 2019 and March 14, 2020. In this case the exposure time was 900 s and the obtained spectra reached a typical signal-to-noise ratio per resolution element of $\sim$80.

In both cases we used the simultaneous calibration technique by taking a spectrum of a Thorium-Argon lamp with the comparison fiber to trace the instrumental drift during the science exposure.

\subsubsection{CHIRON}
The CHIRON instrument \citep{chiron} is a high-resolution and fiber-fed spectrograph mounted on the 1.5 m Smarts telescope at CTIO, Chile. We collected a total of 13 spectra of \stnameA\ with CHIRON, between March 8 and April 10, 2019. For this object we used the image slicer mode (R=80,000), with exposure times between 750 and 1200 s, leading to a mean signal-to-noise per pixel of 33. From this dataset, we computed precision radial velocities following the method described in \citet{wang:2018}, \citet{jones:2019} and \citet{jordan:2020}. We achieve a mean radial velocity precision of 9 m\,s$^{-1}$.

\subsubsection{TRES}
The Tillinghast Reflector Echelle Spectrograph \citep[TRES;][]{furesz:2008}\footnote{\url{http://www.sao.arizona.edu/html/FLWO/60/TRES/GABORthesis.pdf}} is a R=44,000 fiber fed instrument mounted on the 1.5 m Tillinghast Reflector at the Fred L. Whipple Observatory (FLWO) on Mt. Hopkins, AZ. TRES was used to obtain 14 spectra of \stnameB\ between October 7, 2019 and January 28, 2020. A full description of the reduction pipeline and radial velocity extraction process can be seen in \citet{Buchhave:2010}. We deviate from this methodology in the creation of the reference template used for the cross-correlation. We created a high signal-to-noise template spectrum by shifting and median-combining all the spectra, and cross-correlating each observed spectrum against this template to determine the final radial velocities. 

The TRES spectra of \stnameB\ were analyzed using the Stellar Parameter Classification (SPC) package \citep{Buchhave:2012}. From this analysis, we estimated the effective temperature, metallicity, surface gravity, and rotational velocity of \stnameB\ to be: $\ensuremath{T_{\rm eff}}$ = 6048 $\pm$ 50 K, $\ensuremath{\log g_{\star}}$ = 4.32 $\pm$ 0.11 dex, $\ensuremath{\left[{\rm Fe}/{\rm H}\right]}$ = +0.32 $\pm$ 0.08 dex, and $v\sin{i}$ = 8.2 $\pm$ 0.5 km s$^{-1}$.

\subsubsection{CORALIE}
CORALIE is a high resolution (R=60,000) fiber-fed spectrograph mounted on the 1.2 m Swiss Euler telescope at La Silla Observatory, Chile. CORALIE is a stabilized instrument with a comparison fiber to trace the instrumental variations during scientific exposures. We obtained 9 CORALIE spectra of \stnameA\ between March 1, 2019 and April 4, 2019 using a Fabry-Perot as wavelength comparison source. The CORALIE data were processed with its standard data reduction software, where radial velocities and line bisector spans are computed via cross-correlation with a G2 binary mask. In an exposure time of 1200 - 1800 s we obtain signal-to-noise ratio per resolution element of about 30 in individual spectra, corresponding to a final radial velocity uncertainty of $\sim$10 m s$^{-1}$.

\subsubsection{{\textsc{\textsc{Minerva}}}-Australis}
{\textsc{\textsc{Minerva}}}-Australis is an array of four PlaneWave CDK700 telescopes which can be simultaneously fiber-fed to a single KiwiSpec R4-100 high-resolution (R=80,000) spectrograph \citep{barnes:2012,2019PASP..131k5003A,addison:2020}. \stnameA\ was monitored by {\textsc{\textsc{Minerva}}}-Australis using one and/or two telescopes in the array (\textsc{\textsc{Minerva}3} and \textsc{\textsc{Minerva}4}) between March 1, 2019 and May 23, 2019, obtaining 54 spectra in the process over 22 different epochs. 
Radial velocities for the observations are derived for each telescope by cross-correlation, where the template being matched is the mean spectrum of each telescope. The instrumental variations are corrected by using simultaneous Thorium-Argon arc lamp observations. Radial velocities computed from different \textsc{\textsc{Minerva}} telescopes are modeled in Section \ref{sec:glob} as originating from independent instruments.


\subsubsection{NRES}
 Las Cumbres Observatory's \citep{brown:2013:lcogt} Network of Robotic Echelle Spectrographs \citep[NRES,][]{siverd:2018} is a global array of echelle spectrographs mounted on 1~m telescopes, with a resolving power of R$\approx 53,000$. TOI-481 was
observed with the NRES node located at the South African Astronomical
Observatory, for 9 nights between March and April, 2019. At each observing
epoch, two or three consecutive exposures were obtained with a total
nightly exposure time of 3600 s. Overall, 21 spectra were obtained,
with an individual signal-to-noise ratio per resolution element larger than 30.

A SpecMatch \citep{yee17} analysis was performed on the NRES spectra and yielded $T_{\rm eff} = 5730 \pm 100{\rm \, K}$, $\log{g} = 3.9 \pm 0.1\ {\rm dex}$, $[{\rm Fe/H}]=+0.34 \pm 0.06\ {\rm dex}$ and $v \sin {i} \lesssim\ 2\ {\rm km \, s^{-1}}$. The radial velocity of each exposure was derived via cross-correlation with a PHOENIX template \citep{husser:2013} with $T_{\rm eff} = 5700\ {\rm K}$, $\log{g} = 4.0$ dex, [Fe/H] = +0.5 dex, and $v \sin {i}= 2\ {\rm km \, s^{-1}}$. Systematic drifts were corrected per order \citep[e.g.,][]{engel17}.


\begin{deluxetable}{lrrrr}
\tablewidth{0pc}
\tablecaption{
    Radial velocity measurements for \stnameA\ and \stnameB. \label{tab:rvs}
}
\tablehead{
    \colhead{ID} &
    \colhead{BJD} & 
    \colhead{RV} & 
    \colhead{\ensuremath{\sigma_{\rm RV}}} &
    \colhead{Instrument} \\
    \colhead{} &    
    \colhead{-2450000} & 
    \colhead{(m s$^{-1}$)} & 
    \colhead{(m s$^{-1}$)} &
    \colhead{} \\    
}
\startdata
TOI-481 & 8543.59063 & 37723.80 & 5.40 & FEROS \\
TOI-481 & 8544.41789 & 37227.70 & 41.69 & NRES \\
TOI-481 & 8544.43251 & 37253.27 & 143.99 & NRES \\
TOI-481 & 8544.44711 & 37203.45 & 123.13 & NRES \\
TOI-481 & 8544.69135 & 37672.30 & 9.40 & CORALIE \\
TOI-481 & 8544.69945 & 37690.60 & 9.40 & FEROS \\
TOI-481 & 8545.41809 & 37032.61 & 120.31 & NRES \\
TOI-481 & 8545.43966 & 37015.83 & 72.10 & NRES \\
TOI-481 & 8546.69284 & 37714.30 & 6.30 & FEROS \\
TOI-481 & 8548.68585 & 37823.70 & 5.50 & FEROS \\
\enddata
\tablecomments{
    This table is available in a machine-readable form in the online
    journal.  A portion is shown here for guidance regarding its form
    and content.
}
\end{deluxetable}

\section{Analysis} 
\label{sec:anal}

\subsection{Properties of the host star} 
\label{sec:star}
We used the co-added FEROS spectra of \stnameA\ and \stnameB\ to obtain their respective atmospheric parameters. They were obtained using the \zaspe\ code \citep{brahm:2016:zaspe}. \zaspe\ works by comparison via $\chi^2$ minimization of the observed spectrum with a grid of synthetic models generated from the ATLAS9 model atmospheres \citep{atlas9b}. The evaluation is performed in a subset of spectral regions that are most sensitive to changes in the atmospheric parameters. The errors on the atmospheric parameters are computed through Monte Carlo simulations where the depth of the absorption lines of the synthetic models are randomly perturbed to account for the systematic model mismatch.
For \stnameA\ we obtain an effective temperature of \teff\ = \teffA\ K, a surface gravity of \logg\ = \loggA\ dex, a metallicity of \feh\ = \fehA\ dex, and a projected rotational velocity of \vsini\ = \vsiniA\ \kms. On the other hand, for \stnameB\ we found the following set of atmospheric parameters: \teff\ = \teffB\ K, \logg\ = \loggB\ dex, \feh\ = \fehB\ dex, and \vsini\ = \vsiniB\ \kms.

For estimating the stellar physical parameters we followed the same procedure presented in \citet{hd1397}. Briefly, we use the PARSEC isochrones \citep{bressan:2012}  containing the Gaia (G, G$_{BP}$, G$_{RP}$) and 2MASS absolute magnitudes for a given set of stellar mass, age and metallicity. We then use the spectroscopic temperature, the observed magnitudes and the Gaia parallax as data to estimate the stellar mass and the age of each system through a Monte Carlo Markov Chain (MCMC) exploration of the parameter space. We fix the metallicity of the isochrones to the value obtained with \zaspe. With this procedure for \stnameA\ we obtained a mass of \mstar\ = \mstA\ \msun,
a stellar radius of \rstar\ = \rstA\ \rsun, and an age of \ageA\ Gyr. These parameters indicate that \stnameA\ is in the final stages of its main sequence lifetime, about to exhaust the hydrogen in its core.
In the case of \stnameB\ we obtained a mass of \mstar\ = \mstB\ \msun,
a stellar radius of \rstar\ = \rstB\ \rsun, and an age of \ageB\ Gyr. \stnameB\ is therefore a metal rich main sequence F-type star. We stress that the uncertainties reported for the stellar physical parameters are internal, and do not account for possible systematic errors associated to the theoretical isochrones. 

\begin{figure}
\plotone{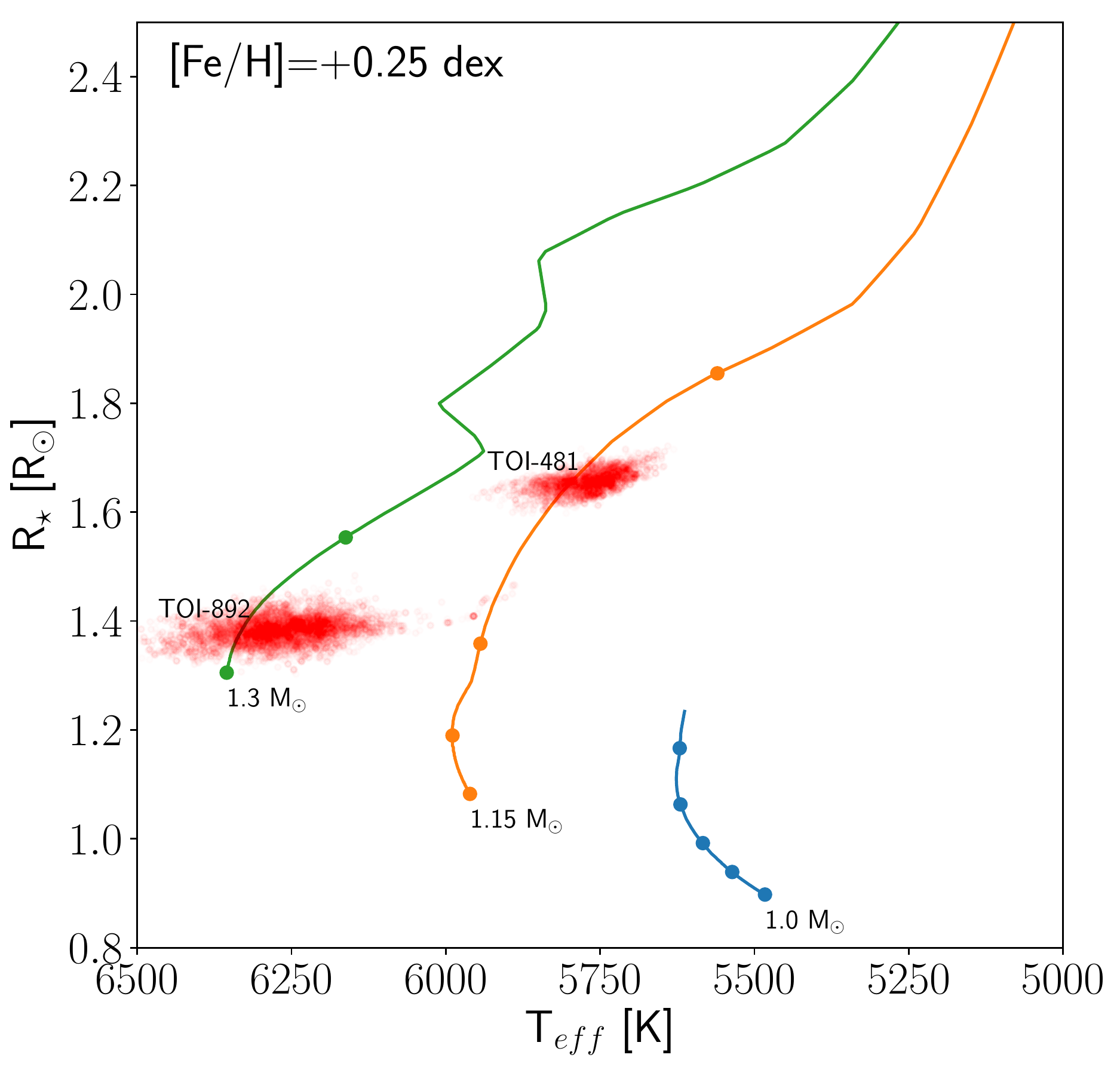}
\caption{Posterior distributions for the effective temperature and stellar radius of \stnameA\ and \stnameB\ (red clumps). The lines represent PARSEC stellar evolutionary tracks for stellar masses of 1 \msun\ (blue), 1.15 \msun\ (orange), and 1.3 \msun\ (green). A metallicity of [Fe/H] = +0.25 is assumed for the three evolutionary tracks, and the circles from bottom to top correspond of the ages of 1, 3, 5, and 7 Gyr. \label{fig:iso}}
\end{figure}

All atmospheric and physical parameters for both stars are presented in Table~\ref{tab:stprops} along with their photometric magnitudes and other observable properties. Additionally, Figure \ref{fig:iso} shows how the distributions for the stellar radius and effective temperature compared to the PARSEC stellar evolutionary models.

We also applied the routines presented in \citet{stassun2018a,stassun:2018} to obtain an independent set of stellar parameters for \stnameA\ and \stnameB. Here we used the Gaia DR2 parallax, along with the \textit{BVgri} magnitudes from APASS, the JHK$_S$ magnitudes from 2MASS, the W1--W4 magnitudes from WISE, and the $G,G_{BP},G_{RP}$ magnitudes from Gaia, to perform a spectral energy distribution fit. This method allow us to determine the stellar radius, metallicity, effective temperature, and surface gravity. All parameters obtained through this method are consistent at one sigma to those listed in Table \ref{tab:stprops}.

\begin{deluxetable*}{lrrc}[b!]
\tablecaption{Stellar properties of TOI-481 and TOI-892.  \label{tab:stprops}}
\tablecolumns{4}
\tablewidth{0pt}
\tablehead{
\colhead{Parameter} &
\colhead{TOI-481} &
\colhead{TOI-892} &
\colhead{Reference} \\
}
\startdata
Names \dotfill   &   &   &   \\
 &  	TIC 339672028 & TIC 66561343 & TICv8 \\
 & 2MASS J07220299-5723054 & UCAC4 394-009979 & 2MASS  \\
 & TYC 8559-00623-1 & TYC 5351-00283-1 & TYCHO  \\
RA \dotfill (J2015.5) &  07h22m03.04s & 05h46m57.17s & TICv8\\
DEC \dotfill (J2015.5) & -57d23m05.99a & -11d14m07.22s & TICv8\\
pm$^{\rm RA}$ \hfill (mas yr$^{-1}$) & 25.68 $\pm$ 0.06 & -0.14 $\pm$ 0.09 & Gaia DR2\\
pm$^{\rm DEC}$ \dotfill (mas yr$^{-1}$) & -25.38 $\pm$ 0.08 & 5.97 $\pm$ 0.10 & Gaia DR2\\
$\pi$ \dotfill (mas)& 5.55 $\pm$ 0.03 & 2.91 $\pm$ 0.04 & Gaia DR2 \\ 
\hline
T \dotfill (mag) & 9.393 $\pm$ 0.006 & 10.974 $\pm$ 0.030 & TICv8\\
B  \dotfill (mag) & 10.68 $\pm$ 0.05 & 12.06 $\pm$ 0.04 & APASS\tablenotemark{a}\\
V  \dotfill (mag) & 10.04 $\pm$ 0.02 & 11.45 $\pm$ 0.02 & APASS\\
G  \dotfill (mag) & 9.846 $\pm$ 0.002 & 11.343 $\pm$ 0.002 & Gaia DR2\tablenotemark{b}\\
G$_{BP}$  \dotfill (mag) & 10.219 $\pm$ 0.005 & 11.643 $\pm$ 0.005 & Gaia DR2\\
G$_{RP}$  \dotfill (mag) & 9.354 $\pm$ 0.004 & 10.907 $\pm$ 0.003 & Gaia DR2\\
J  \dotfill (mag) &  8.80 $\pm$ 0.02 & 10.46 $\pm$ 0.03 & 2MASS\tablenotemark{c}\\
H  \dotfill (mag) &  8.48 $\pm$ 0.04 & 10.19 $\pm$ 0.02 & 2MASS\\
K$_s$  \dotfill (mag) & 8.443 $\pm$ 0.02 & 10.11 $\pm$ 0.02 & 2MASS\\
\hline
\teff  \dotfill (K) & \teffA & \teffB & This work\\
\logg \dotfill (dex) & \loggA & \loggB & This work\\
\feh \dotfill (dex) & \fehA & \fehB & This work\\
\vsini \dotfill (km s$^{-1}$) & \vsiniA & \vsiniB & This work\\
\mstar \dotfill (\msun) & \mstA & \mstB & This work\\
\rstar \dotfill (\rsun) & \rstA & \rstB & This work\\
L$_{\star}$ \dotfill (L$_{\odot}$) & \lstA & \lstB & This work\\
A$_{V}$ \dotfill (mag) & \avA & \avB & This work\\
Age \dotfill (Gyr) & \ageA & \ageB & This work\\
$\rho_\star$ \dotfill (g cm$^{-3}$) & \rhostA & \rhostB  & This work\\
\enddata
\tablecomments{$^a$\citet{apass},$^b$\citet{gaia:dr2},$^c$\citet{2mass}}.
\end{deluxetable*}

\subsection{Radial Velocities}
\begin{figure}
\epsscale{1.1}
\plotone{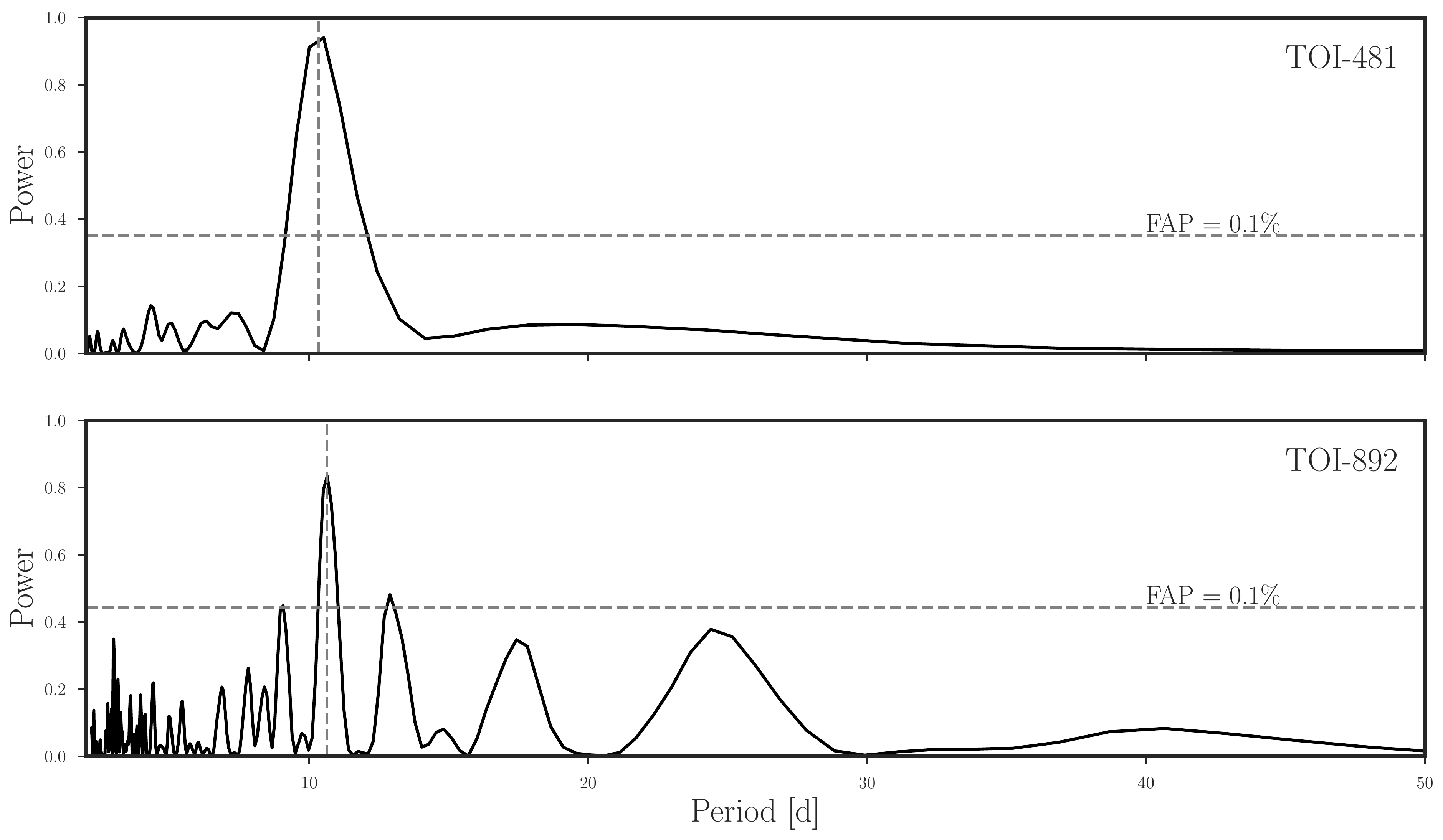}
\caption{ Lomb scargle periodograms for the radial velocity time series of \stnameA\ (top panel) and \stnameB\ (bottom panel). The horizontal dashed line corresponds to the 0.1\% false alarm probability. The vertical dashed line corresponds to the period of the transiting candidates.
 \label{fig:perio}}
\end{figure}

We analysed the radial velocity time series of both systems to identify variations consistent with the presence of orbiting planets having the periodicity of the transiting candidates. We computed the Generalized Lomb-Scargle periodograms by combining the radial velocities of different instruments for each system. The periodograms are presented in Figure \ref{fig:perio} which confirms that both radial velocity sets have significant periodicity at the orbital period of the transiting candidate. The semi-amplitude of these periodic radial velocity variations is consistent with that of giant planets in moderately close-in orbits (K$\sim$100 m s$^{-1}$).

In order to further confirm that the radial velocity signals are produced by orbiting planets, we analysed the degree of correlation between the radial velocities and line bisector span measurements. We computed the Pearson correlation coefficient with errors through bootstrap finding $\rho_P=0.22 \pm 0.19$ and $\rho_P=-0.19 \pm 0.25$, for \plnameA\ and \plnameB, respectively. Therefore, the absence of significant correlation between radial velocities and line bisector span measurements further supports the hypothesis that the radial velocity variations for both systems are produced by the gravitational pull of transiting giant planets.

\subsection{Global Modeling} \label{sec:glob}
The global modeling of the photometric data and radial velocities for the \stnameA\ and \stnameB\ systems was performed with the \texttt{juliet} package \citep{juliet}.
This package can use either MultiNest \citep{MultiNest} through the PyMultiNest package 
\citep{PyMultiNest} or \texttt{dynesty} \citep{dynesty} to perform the posterior sampling via nested sampling algorithms, in order to also compute model comparison through Bayesian model evidences.
\texttt{juliet} uses \texttt{batman} \citep{kreidberg:2015:bat} to model the photometric transits, while radial velocities variations are modelled with the \texttt{radvel} package \citep{fulton:2018}.

\begin{figure*}
\epsscale{1.1}
\plottwo{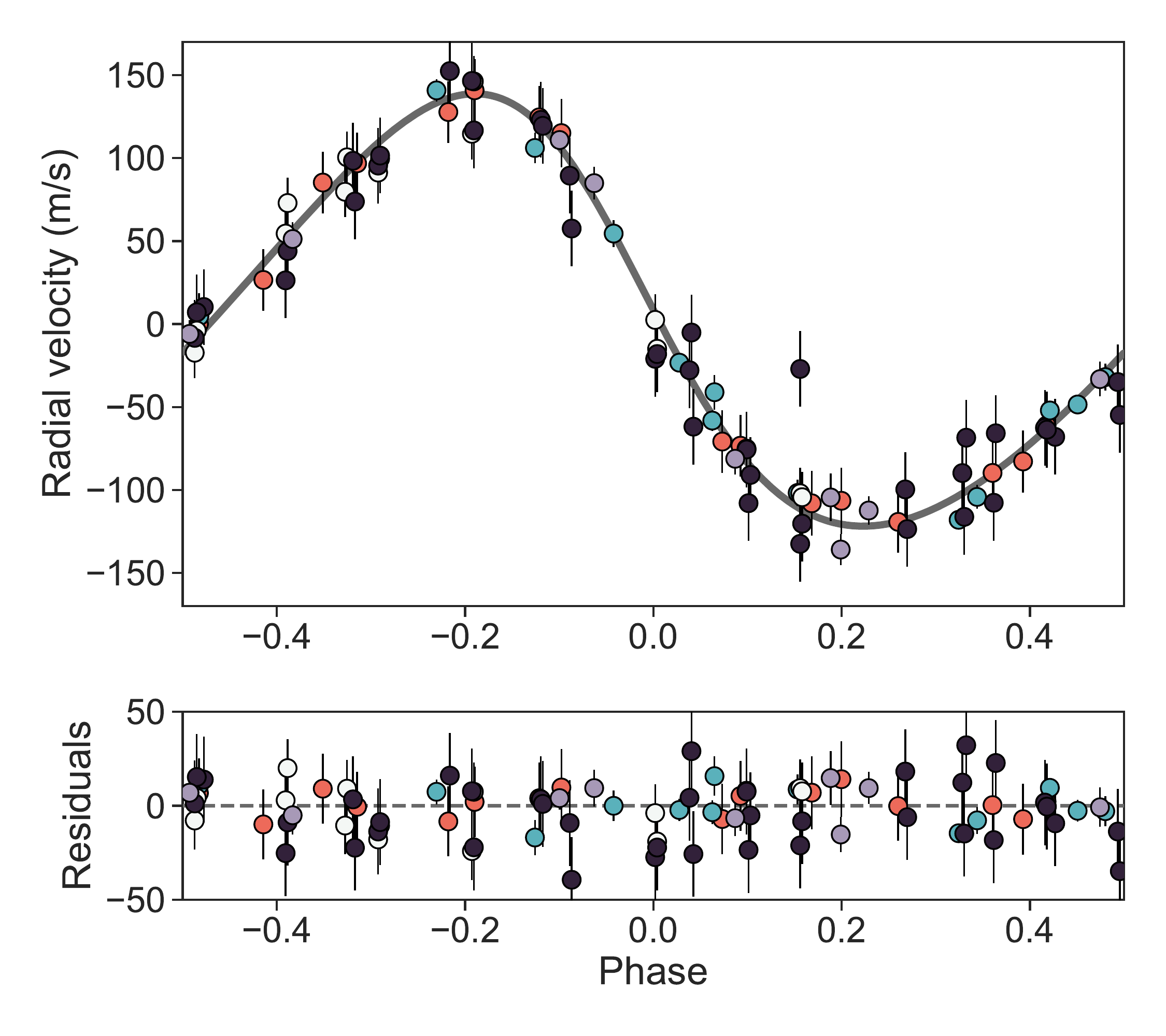}{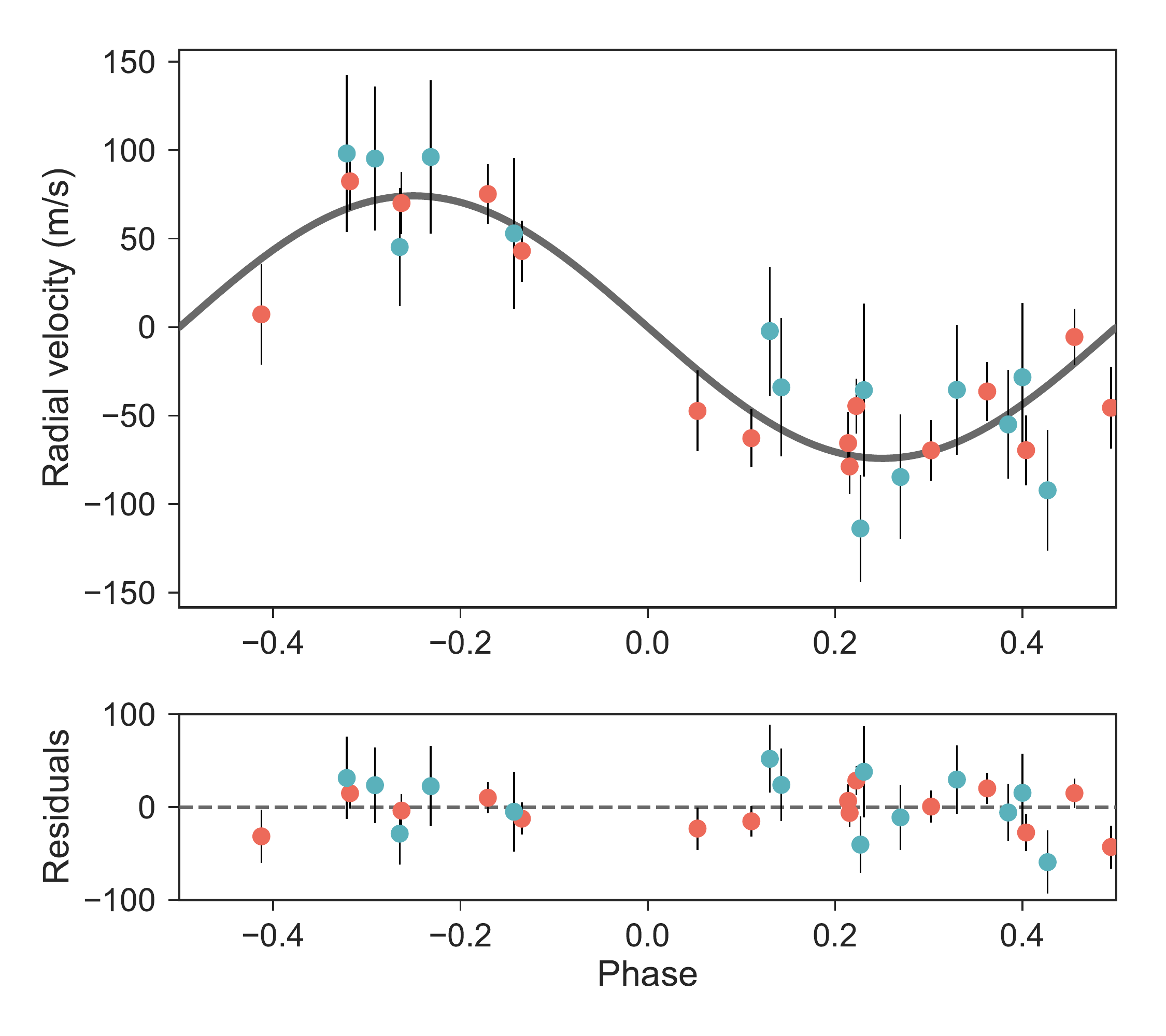}
\caption{The left panel shows the radial velocities of \plnameA\ as a function of the orbital phase. The black line represents the model generated from the posterior distributions obtained in Section \ref{sec:anal}. The errorbars include the jitter term obtained from the global analysis. The different colors represent the different instruments that were used, namely: FEROS (orange), CHIRON (blue), CORALIE (gray), MINERVA-3 (white), and MINERVA-4 (black). The right panel shows the same but for the FEROS (orange) and TRES (green) velocities of \plnameB.
 \label{fig:rvs_fold}}
\end{figure*}

The parameters that were considered for modelling each system are described in the following paragraphs. The photometric and phase folded radial velocity models that were obtained with this process are presented in Figures \ref{fig:lcs_fold} and \ref{fig:rvs_fold}, respectively, along with the corresponding observations.

\subsubsection{Global modelling of the \stnameA\ system}
For the \stnameA\ system, we ran \texttt{juliet} fits using \texttt{dynesty}, as the number of free parameters (54) needed to account for the global fit is relatively large. In this global 
fit, we used the \cite{espinoza:2018:r1r2} parametrization to fit for the planet-to-star radius ratio and the impact parameter. 
On top of this, we used a prior on the 
stellar density given by our analysis of the stellar properties in the previous subsection.

For the \textit{TESS} photometry we used a Mat\'ern 3/2 kernel implemented via \texttt{celerite} \citep{celerite} to model systematic trends 
with individual hyperparameters for the amplitude and time-scale of the process for each \textit{TESS} Sector. 
For the limb-darkening, we assumed a quadratic limb-darkening law 
parametrized using the non-informative sampling scheme outlined in \cite{Kipping:LDs}. 
For the short-cadence photometry 
we assumed unitary dilution factors, as these 
are already accounted for by the Pre-search Data Conditioning algorithm. For the long-cadence 
photometry of Sector 3 that was extracted with our own aperture photometry, we considered a dilution factor with a prior between 0.95 and 1. 
We included photometric jitter terms for each Sector, but the measured jitter was consistent with zero for all but Sector 3. We therefore only fit for photometric jitter in Sector 3 data.

For the CHAT and NGTS photometry, we found no evidence of obvious systematic trends and thus decided to model those datasets as having white-gaussian noise.
We used a linear limb-darkening law for both ground-based light curves and we fit for a dilution factor term in the case of the NGTS photometry. 

Finally, for the radial velocities we used simple white-gaussian 
noise models, where each instrument has its own systemic velocity and jitter term. 
Fits using both eccentric and non-eccentric orbits were performed, 
with the eccentric model being drastically preferred over the non-eccentric model ($ln Z>5$ in favor of the eccentric model). The final posterior parameters of the global analysis of \plnameA\ are presented in Table \ref{tab:plpropsA}, along with the prior distributions used for each parameter.

By combining the stellar properties of \stnameA\ with the posterior parameters of the adopted \texttt{juliet} fit, we find that \plnameA\ has a mass of \mpl\ = \mpA\ \mjup, a radius of \rpl\ = \rpA\ \rjup, a time averaged equilibrium temperature \citep{mendez:2017} of \teq\ = \tqA\ K (partial heat distribution of $\beta$=0.5 and bond Albedo $A=1$), and an orbital eccentricity of $e = \eccA$. 

\subsubsection{Global modelling of the \stnameB\ system}
For the \stnameB\ system we ran two \texttt{juliet} fits using PyMultiNest, as the number of free parameters (19) is much smaller in this case; one with the eccentricity and argument of the periastron as free parameters, and another one with those values fixed to 0. In both cases for modeling the \texttt{tesseract} light curve we adopted the quadratic law for the limb darkening, and a gaussian process with a Matern 3/2 Kernel to model the out of transit variations. For the MEarth and CHAT light curves we adopted a linear limb darkening law. For the radial velocities we considered independent zero points and jitter terms for each spectrograph. We found that the joint modeling with zero eccentricity delivers a higher log evidence than the eccentric version, and we adopted the posterior parameters of that model, which are presented in Table \ref{tab:plpropsB} along with the derived planet parameters.

By combining the posterior parameters of the adopted joint fit with the stellar properties of \stnameB\ we find that \plnameB\ has a mass of \mpl\ = \mpB\ \mjup, a radius of \rpl\ = \rpB\ \rjup, and an equilibrium temperature of \teq\ = \tqB\ K. We determine a 98\% upper limit for the orbital eccentricity of \plnameB\ of 0.125.

{\setlength\tabcolsep{3.3pt} 
\LTcapwidth=1.1\columnwidth
\begin{longtable}{lrr}
\caption{Prior and posterior parameters of the global analysis of \plnameA. For the priors, $N(\mu,\sigma)$ stands for a normal distribution with mean $\mu$ and standard deviation $\sigma$, $U(a,b)$ stands for a uniform distribution between $a$ and $b$, and $LU(a,b)$ stands for a log-uniform prior defined between $a$ and $b$.\\
$^{a}$ These parameters correspond to the parametrization presented in \citet{espinoza:2018:r1r2} for sampling physically possible combinations of $b$ and \rpl/\rstar.\\
$^{b}$ Time-averaged equilibrium temperature computed according to equation~16 of \citet{mendez:2017}.
\label{tab:plpropsA}}\\
\hline
\hline
\noalign{\smallskip}
Parameter & Prior  & Value \\
\hline
\endfirsthead
\caption{Continued.} \\
\hline
Parameter & Prior  & Value \\
\hline
\endhead
\hline
\endfoot
\noalign{\smallskip}
\noalign{\smallskip}
P \dotfill (days) & $N(10.331,0.1)$  &          \PA \\
T$_0$ \dotfill (BJD)&  $N(2458511.641,0.1)$&  \tcA \\
$\rho_\star$ \dotfill (g cm$^{-3}$)  & $U(0.36,0.01)$ & \aA \\
r1\tablenotemark{a} \dotfill & $U(0,1)$ & \raA \\
r2\tablenotemark{a} \dotfill & $U(0,1)$ & \rbA \\
K \dotfill (m s$^{-1}$) & $U(0,1000)$& \KA \\
$\sqrt{e}\sin{\omega}$ \dotfill & $U(-1,1)$ & \sesinomegaA \\
$\sqrt{e}\cos{\omega}$ \dotfill & $U(-1,1)$ & \secosomegaA \\
q$_1^{\rm TESS}$ \dotfill & $U(0,1)$ & \qaTESSA \\
q$_2^{\rm TESS}$ \dotfill & $U(0,1)$& \qbTESSA \\
q$_1^{\rm CHAT}$ \dotfill & $U(0,1)$ & \qaCHATA \\
q$_1^{\rm NGTS}$ \dotfill & $U(0,1)$ & \qaNGTSA \\
$\sigma_w^{\rm TESS-S3}$ \dotfill (ppm) & $LU(10^{-2},10^{3})$ & $459^{+8}_{-9}$ \\
$\sigma_w^{\rm CHAT}$ \dotfill (ppm) & $LU(10^{-2},10^{4})$ & \sigmaCHATA \\
$mflux^{\rm TESS-S3}$  \dotfill & $N(0,0.1)$ &  $-0.00001^{+0.00005}_{-0.00005}$\\
$mflux^{\rm TESS-S6}$  \dotfill & $N(0,0.1)$ &  $-0.0000006^{+0.00005}_{-0.00005}$\\
$mflux^{\rm TESS-S7}$  \dotfill & $N(0,0.1)$ &  $0.00001^{+0.00004}_{-0.00004}$\\
$mflux^{\rm TESS-S9}$  \dotfill & $N(0,0.1)$ &  $-0.00004^{+0.00003}_{-0.00002}$\\
$mflux^{\rm TESS-S10}$ \dotfill & $N(0,0.1)$ &  $-0.00003^{+0.00003}_{-0.00003}$\\
$mflux^{\rm TESS-S13}$ \dotfill & $N(0,0.1)$ &  $0.001^{+0.004}_{-0.001}$\\
$mflux^{\rm CHAT}$     \dotfill & $N(0,0.1)$ &  $-0.0019^{+0.0001}_{-0.0002}$\\
$mflux^{\rm NGTS}$     \dotfill & $N(0,0.1)$ &  $0.00004^{+0.00007}_{-0.00007}$\\
$dilution^{\rm TESS-S3}$  \dotfill & $U(0.95,1.0)$ &  $0.982^{+0.004}_{-0.005}$\\
$dilution^{\rm NGTS}$     \dotfill & $U(0.95,1.0)$ &  $0.995^{+0.003}_{-0.003}$\\
$\gamma_{\rm CHIRON}$  \dotfill (m s$^{-1}$)& $N(0,50)$& \muCHIRONA \\
$\gamma_{\rm CORALIE}$ \dotfill (m s$^{-1}$)& $N(37800,50)$ & \muCORALIEA \\
$\gamma_{\rm FEROS}$ \dotfill (m s$^{-1}$)& $N(37800,50)$ & \muFEROSA \\
$\gamma_{\rm \textsc{Minerva}-3}$ \dotfill (m s$^{-1}$)& $N(0,50)$ & \muAMinervaA \\
$\gamma_{\rm \textsc{Minerva}-4}$ \dotfill (m s$^{-1}$)& $N(0,50)$ & \muBMinervaA \\
$\gamma_{\rm NRES}$ \dotfill (m s$^{-1}$)& $N(0,50)$ & \muNRESA \\
$\sigma_{\rm CHIRON}$  \dotfill (m s$^{-1}$)& $LU(0.01,50)$& \sigmawCHIRONA \\
$\sigma_{\rm FEROS}$ \dotfill (m s$^{-1}$)& $LU(0.01,50)$ & \sigmawFEROSA \\
$\sigma_{\rm \textsc{Minerva}-3}$ \dotfill (m s$^{-1}$)& $LU(0.01,50)$ & \sigmawAMinervaA \\
$\sigma_{\rm \textsc{Minerva}-4}$ \dotfill (m s$^{-1}$)& $LU(0.01,50)$ & \sigmawBMinervaA \\
$\sigma_{\rm NRES}$ \dotfill (m s$^{-1}$)& $LU(0.01,50)$ & \sigmawNRESA \\
$\sigma^{GP}_{\rm TESS-3}$ \dotfill &  $LU(10^{-5},10^{2})$ &  $0.00031^{+0.00003}_{-0.00003}$\\
$\sigma^{GP}_{\rm TESS-6}$ \dotfill &  $LU(10^{-5},10^{2})$ &  $0.00021^{+0.00004}_{-0.00003}$ \\
$\sigma^{GP}_{\rm TESS-7}$ \dotfill &  $LU(10^{-5},10^{2})$ &  $0.00014^{+0.00003}_{-0.00002}$\\
$\sigma^{GP}_{\rm TESS-9}$ \dotfill &  $LU(10^{-5},10^{2})$ &  $0.00018^{+0.00002}_{-0.00002}$\\
$\sigma^{GP}_{\rm TESS-10}$ \dotfill &  $LU(10^{-5},10^{2})$ & $0.00017^{+0.00002}_{-0.00002}$\\
$\sigma^{GP}_{\rm TESS-13}$ \dotfill &  $LU(10^{-5},10^{2})$ & $0.003^{+0.005}_{-0.002}$\\
$\rho^{GP}_{\rm TESS-3}$ \dotfill &  $LU(10^{-2},10^{2})$  & $0.37^{+0.07}_{-0.05}$\\
$\rho^{GP}_{\rm TESS-6}$ \dotfill &  $LU(10^{-2},10^{2})$  & $0.95^{+0.21}_{-0.16}$\\
$\rho^{GP}_{\rm TESS-7}$ \dotfill &  $LU(10^{-2},10^{2})$  & $0.9^{+0.3}_{-0.2}$\\
$\rho^{GP}_{\rm TESS-9}$ \dotfill &  $LU(10^{-2},10^{2})$  & $0.35^{+0.05}_{-0.05}$\\
$\rho^{GP}_{\rm TESS-10}$ \dotfill &  $LU(10^{-2},10^{2})$ & $0.47^{+0.07}_{-0.06}$  \\
$\rho^{GP}_{\rm TESS-13}$ \dotfill &  $LU(10^{-2},10^{2})$ & $21^{+23}_{-11}$  \\
\hline
b & \dotfill  & \bA \\
\rpl/\rstar  & \dotfill  & \pA \\
$e$ & \dotfill  & \eccA \\
$\omega$ (deg) & \dotfill  & \omegaA \\
$i$ (deg) & \dotfill  & \incA \\
\mpl\ (\mjup)& \dotfill & \mpA \\
\rpl\ (\rjup)& \dotfill & \rpA \\
$a$ (AU)    & \dotfill & \smaA \\
\teq (K) \tablenotemark{b}   & \dotfill & \tqA \\ 
\hline
\end{longtable}
}

\begin{deluxetable}{lrr}[b!]
\tablecaption{Same as Table \ref{tab:plpropsB}, but for \plnameB. $J(a,b)$ stands for a Jeffrey's prior defined between $a$ and $b$.\label{tab:plpropsB}}
\tablecolumns{3}
\tablewidth{0pt}
\tablehead{
\colhead{Parameter} &
\colhead{Prior} &
\colhead{Value}
}
\startdata
P (days) & $N(10.6,1.0)$  &          \PB \\
T$_0$ (BJD)&  $N(2458475.7,1.0)$&  \tcB \\
$a$/R$_\star$ & $U(1,300)$ & \aB \\
\rpl/\rstar  & $U(0.0001,1)$ & \pB \\
b & $U(0,1)$ & \bB \\
K (km s$^{-1}$) & $U(0,1)$& \KB \\
q$_1^{\rm TESS}$ & $U(0,1)$ & \qaTESSERACTTESSB \\
q$_2^{\rm TESS}$ & $U(0,1)$& \qbTESSERACTTESSB \\
q$_1^{\rm CHAT}$ & $U(0,1)$ & \qaCHATB \\
q$_1^{\rm MEarth}$ & $U(0,1)$ & \qaMEARTHB \\
$\sigma_w^{\rm TESS}$ (ppm) & $J(10^0,10^3)$ & \sigmawTESSERACTTESSB\\
$\sigma_w^{\rm CHAT}$ (ppm) & $J(10^0,10^3)$ & \sigmawCHATB \\
$\sigma_w^{\rm MEarth}$ (ppm) & $J(10^0,10^3)$ & \sigmawMEARTHB \\
$\gamma_{\rm FEROS}$  (km s$^{-1}$)& $N(42.02,0.010)$& \muFEROSB \\
$\gamma_{\rm TRES}$ (km s$^{-1}$)& $N(0.04.0,0.010)$ & \muTRESB \\
$\sigma_{\rm FEROS}$ (km s$^{-1}$)& $N(0.001,0.1)$ & \sigmawFEROSB \\
$\sigma_{\rm TRES}$  (km s$^{-1}$)& $N(0.001,0.1)$& \sigmawTRESB \\
$\sigma^{GP}_{\rm TESS}$ &  $J(10^{-5},10^{3})$ & \GPsigmaTESSERACTTESSB \\
$\rho^{GP}_{\rm TESS}$ &  $J(10^{-5},10^{3})$ & \GPrhoTESSERACTTESSB \\
\hline
$e$ & & $<0.125$ (98$\%$ confidence)\\
$i$ (deg) &  & 88.2$^{+0.3}_{-0.5}$ \\
$\rho_{\star}$ (Kg m$^{-3}$) & & \rhoB \\
\mpl\ (\mjup)& & \mpB \\
\rpl\ (\rjup)& & \rpB \\
$a$ (AU)    & & \smaB \\
\teq (K)  & & \tqB \\ 
\enddata
\end{deluxetable}

\subsection{Timing of transits and additional photometric signals}
\begin{figure*}
\epsscale{2.3}
\plottwo{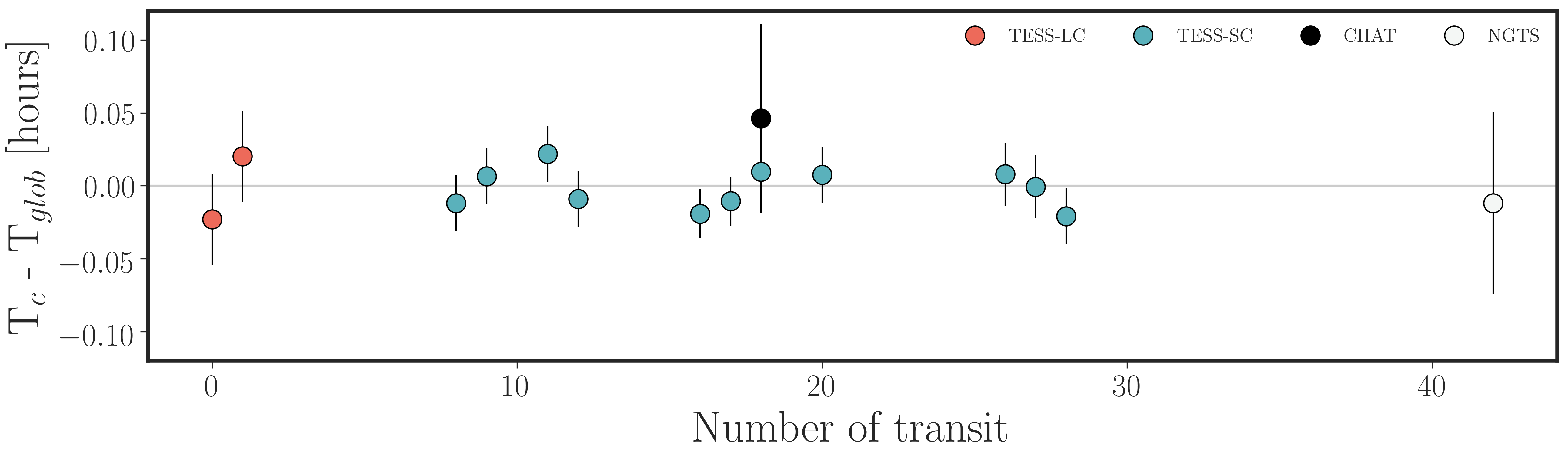}{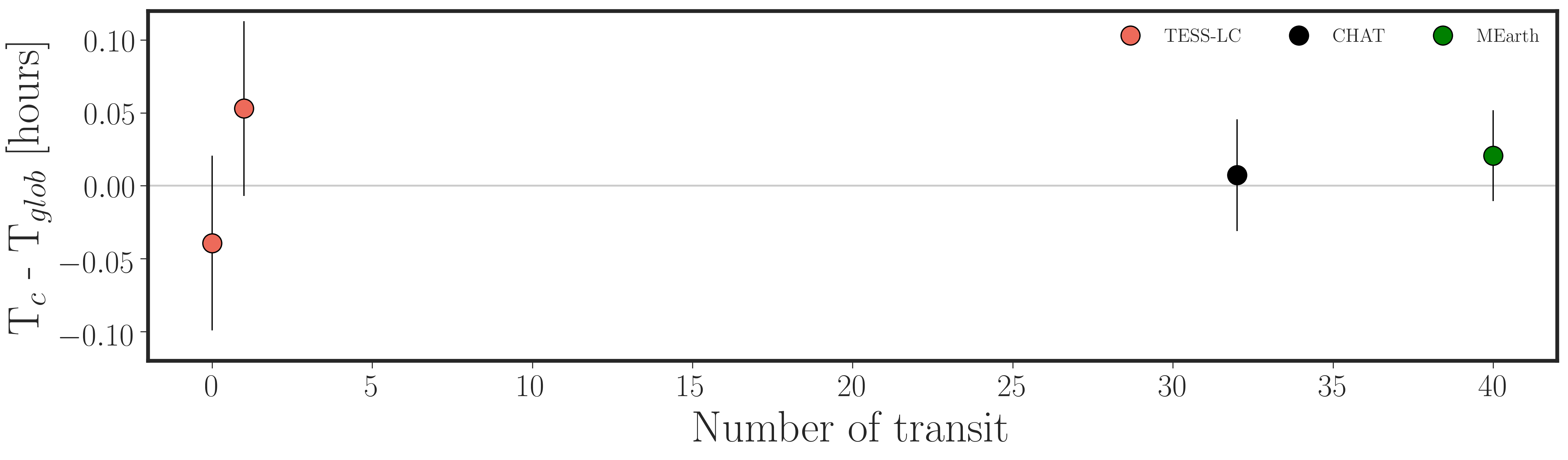}
\caption{Transit timing variations for \plnameA\ (top panel) and \plnameB\ (bottom panel) computed from the \textit{TESS} and ground-based light curves. No significant signal is identified in both cases.
 \label{fig:ttvs}}
\end{figure*}

We searched for variations in the time of transits of \plnameA\ and \plnameB\ that could originate from gravitational interactions with other planets in each system. For this procedure we performed independent \texttt{juliet} runs for each one of the transits of the \textit{TESS} and follow-up light curves. We fix most of the parameters to those obtained in the global analysis, but allowed the time of transit and the transit depth to vary. The transit timing variations for \plnameA\ and \plnameB\ are displayed in Figure \ref{fig:ttvs}. No significant variations in the timing of transits are identified for both systems.

We also searched for additional transiting candidates in the \textit{TESS} data of both systems by masking out the transits of \plnameA\ and \plnameB, and running the box least squares \citep{BLS} algorithm. No significant signals were identified.

\section{Discussion} \label{sec:disc}
\begin{figure*}
\epsscale{1.2}
\plotone{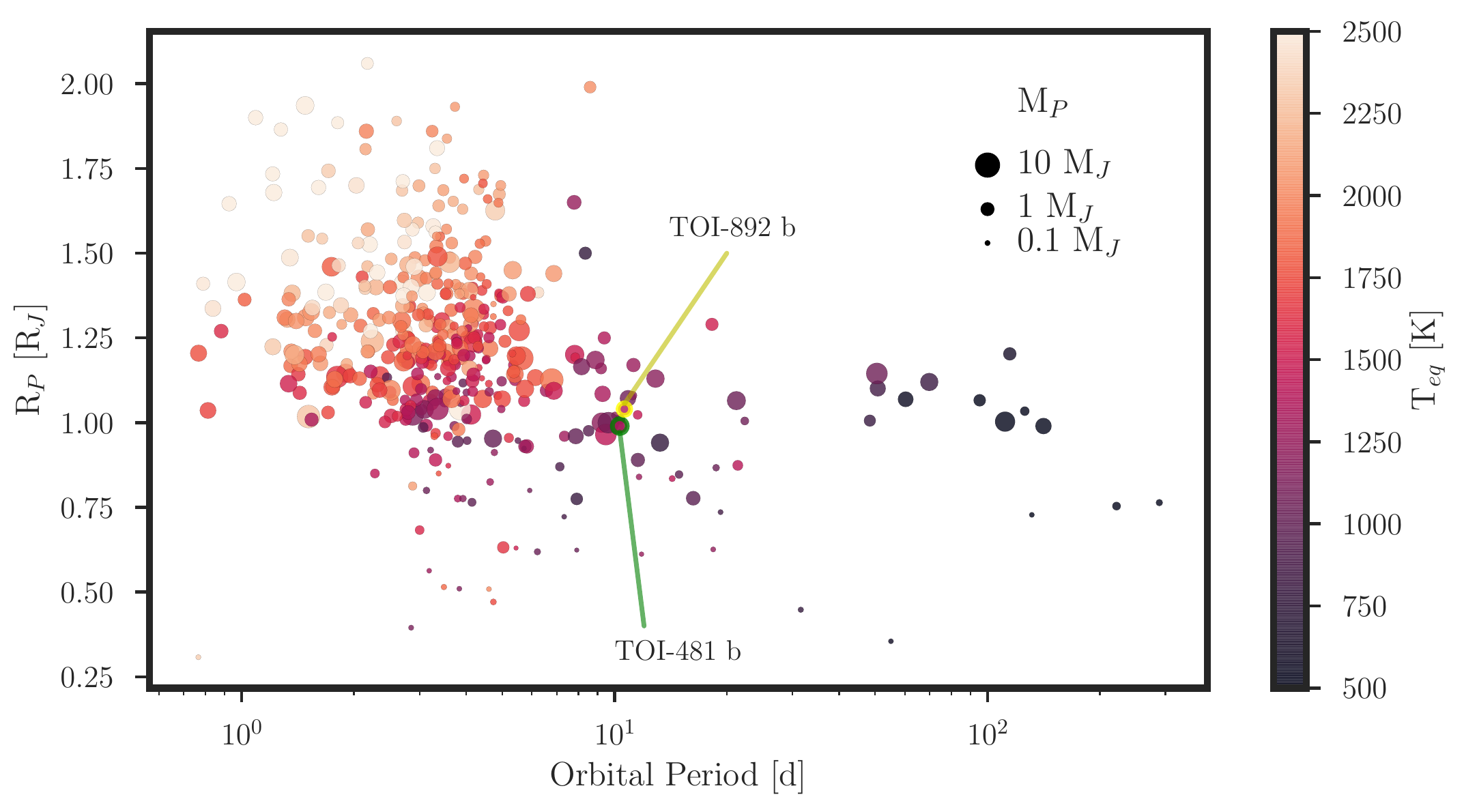}
\caption{Mass--Period diagram for the population of transiting giant planets (\mpl\ $>$ 0.1 \mjup) having masses and radii measured with a precision of 20\% or better. The points are color coded by equilibrium temperature and the pint size scales with the planet mass.
 \label{fig:PR}}
\end{figure*}

\plnameA\ and \plnameB\ are both compared in Figure \ref{fig:PR} with the population of well characterized transiting giant planets (M$_P>0.1$ M$_J$) in the planetary radius versus orbital period space. Both planets join the population of moderately long period (P $>$ 10 days) giant planets, which has just recently started to see an increase in the number of detected systems. In terms of physical and orbital properties \plnameA\ is similar to WASP-134~b \citep[\mpl = 1.41 $\pm$ 0.08 \mjup, \rpl = 0.99 $\pm$ 0.06 \rjup, P = 10.2 days,][]{anderson:2018}. On the other hand, \plnameB\ shares similar properties with WASP-185~b \citep{hellier:2019}, that has a mass of \mpl\ = 0.98 $\pm$ 0.06 \mjup, a radius of \rpl\ = 1.25 $\pm$ 0.08 \rjup\ and an orbital period of P = 9.4 days, but as opposed to \plnameB\ has a significantly eccentric orbit (e=0.23 $\pm$ 0.04).

Despite having periods longer than 10 days, both \plnameA\ and \plnameB\ have moderately high equilibrium temperatures, due to the high luminosity of their host stars. Their equilibrium temperatures are somewhat higher than 1000 K, and therefore these planets are just in the region where the inflation mechanism of hot Jupiters starts to have an impact on the structure of the planet \citep{demory:2011,laughlin:2011}. The measured radius for \plnameB\ is in good agreement with the mean radius of other hot Jupiters having similar insolation levels, which is of 1.1 $\pm$ 0.1 \rjup\ for 1300 $< \teq <$ 1500. The radius of \plnameA\, while still consistent with this distribution, is significantly more compact than that of \plnameB. In this context it is important to note that \stnameA\ is in the final stages of the main sequence evolution and has started to receive increased levels of irradiation during the last Gyr of evolution. Its non-inflated radius can be linked with a shallow level deposition of the stellar energy into the planet interior during main sequence evolution as argued by \citet{komacek:2020}, which is not enough to re-inflate the planet even at temperatures higher than 1000 K. If warm Jupiters are efficiently re-inflated during post-main sequence evolution, as some recent studies have proposed \citep{grunblatt:2016,jones:2018,grunblatt:2017}, then some other mechanism should operate to allow the deposition of energy deeper in the planet interior. 

Both systems are well suited objects to perform Rossiter-McLaughlin observations \citep{rossiter:24,mclaughlin:24} for measuring their stellar obliquities \citep[e.g.,][]{triaud:2010}. Given the properties of each system, we expect them to have Rossiter-McLaughlin signals with semi-amplitudes of 15 m s$^{-1}$ and 47 m s$^{-1}$, for \plnameA\ and \plnameB, respectively, for aligned orbits, which can be measured with typical facilities having a stabilized high resolution echelle spectrograph. Spin-orbit angles of giant planets with orbital separations larger than $\gtrsim$0.1 AU are expected to be particularly useful for constraining migration scenarios, because at these moderately long orbital distances, tidal interactions are not supposed to be strong enough for realigning the rotation of the outer layers of the star with the orbital plane \citep{albrecht:2012,dawson:2014}. The low eccentricities of the orbits of \plnameA\ and \plnameB\, and the absence of close planet companions, based on the radial velocity and photometric data, points to interactions with the protoplanetary disc as the most probable migration scenario for these systems \citep{dong:2014}.

\acknowledgments

We acknowledge the use of public \tess\ Alert data from pipelines at the \tess\ Science Office and at the \tess\ Science Processing Operations Center.
This research has made use of the Exoplanet Follow-up Observation Program website, which is operated by the California Institute of Technology, under contract with the National Aeronautics and Space Administration under the Exoplanet Exploration Program.
This paper includes data collected by the \tess\ mission, which are publicly available from the Mikulski Archive for Space Telescopes (MAST).
Resources supporting this work were provided by the NASA High-End Computing (HEC) Program through the NASA Advanced Supercomputing (NAS) Division at Ames Research Center for the production of the SPOC data products.
R.B.\ acknowledges support from FONDECYT Post-doctoral Fellowship Project 3180246, and from 
ANID – Millennium Science Initiative – ICN12\_009.
A.J.\ acknowledges support from FONDECYT project 1171208, and from ANID – Millennium Science Initiative – ICN12\_009.

JIV acknowledges support of CONICYT-PFCHA/Doctorado Nacional-21191829
This work was supported by the DFG Research Unit FOR2544 ``Blue Planets around Red Stars", project no. RE 2694/4-1.
We thank Paul Eigenthaler, Angela Hempel, Maren Hempel, Sam Kim and R\'egis Lachaume for
their technical assistance during the observations at the MPG 2.2 m Telescope.
We thank the Swiss National Science Foundation (SNSF) and the Geneva University for their continuous support to our planet search programs. This work has been in particular carried out in the frame of the National Centre for Competence in Research PlanetS supported by the Swiss National Science Foundation (SNSF).
\textsc{Minerva}-Australis is supported by Australian Research Council LIEF Grant LE160100001, Discovery Grant DP180100972, Mount Cuba Astronomical Foundation, and institutional partners University of Southern Queensland, UNSW Sydney, MIT, Nanjing University, George Mason University, University of Louisville, University of California Riverside, University of Florida, and The University of Texas at Austin.
This research was supported by Grant No. 2016069 of the United States-Israel Binational Science Foundation (BSF)
The MEarth Team gratefully acknowledges funding from the David and Lucile Packard Fellowship for Science and Engineering (awarded to D.C.). This material is based upon work supported by the National Science Foundation under grants AST-0807690, AST-1109468, AST-1004488 (Alan T. Waterman Award), and AST- 1616624, and upon work supported by the National Aeronautics and Space Administration under Grant No. 80NSSC18K0476 issued through the XRP Program. This work is made possible by a grant from the John Templeton Foundation. The opinions expressed in this publication are those of the authors and do not necessarily reflect the views of the John Templeton Foundation.
We respectfully acknowledge the traditional custodians of all lands throughout Australia, and recognise their continued cultural and spiritual connection to the land, waterways, cosmos, and community. We pay our deepest respects to all Elders, ancestors and descendants of the Giabal, Jarowair, and Kambuwal nations, upon whose lands the \textsc{Minerva}-Australis facility at Mt Kent is situated.




\software{\texttt{juliet} \citep{juliet},
          \ceres \citep{brahm:2017:ceres,jordan:2014},
          \zaspe \citep{brahm:2016:zaspe,brahm:2015},
          \texttt{radvel} \citep{fulton:2018}
          \texttt{emcee} \citep{emcee:2013},
          MultiNest \citep{MultiNest},
          \texttt{batman} \citep{kreidberg:2015:bat},
          \texttt{SPC} \citep{Buchhave:2012},
          \texttt{SpecMatch} \citep{yee17}
          }
          
\facilities{
 	{\it Astrometry}:
 	Gaia \citep{gaia:2016:dr1,gaia:dr2}.
 	{\it Imaging}:
    SOAR~(HRCam; \citealt{SOAR}).
 	{\it Spectroscopy}:
	CTIO1.5m~(CHIRON; \citealt{chiron}),
    MPG2.2m~(FEROS; \citealt{kaufer:99}),
    Euler1.2m~(CORALIE; \citealt{mayor:2003}),
    Tillinghast1.5m~(TRES) \citep{furesz:2008},
    {\textsc{\textsc{Minerva}}}-Australis \citep{2019PASP..131k5003A},
    NRES \citep{siverd:2018},
 	{\it Photometry}:
 	CHAT:0.7m,
    MEarth-South \citep{irwin:2015},
	NGTS \citep{wheatley:2018},
 	TESS \citep{tess}.
}

\bibliography{k2clbib}




\end{document}